\documentclass[a4paper]{article}

\usepackage{empheq}
\usepackage{float} 
\usepackage{microtype}
\usepackage{enumitem}
\usepackage{booktabs}
\usepackage{hyperref}

\usepackage[authoryear,numbers]{natbib}
\usepackage{amssymb}

\usepackage{tabularx}

\newcommand{\bsigma}{\boldsymbol{\bar{\sigma}}}
\newcommand{\bchi}{\boldsymbol{\bar{\chi}}}
\newcommand{\byield}{\bar{\sigma}_\mathrm{Y}}
\newcommand{\bN}{\bar{\boldsymbol {N}}}

\title{\textbf{Coupling of the phase field approach to the Armstrong-Frederick model for the simulation of ductile damage under cyclic load}}
\author{\textbf{Serhat Aygün$^1$, Tillmann Wiegold$^2$ and Sandra Klinge$^{1,}$\thanks{Correspondence: sandra.klinge@tu-berlin.de}}  \\ \\
    {\normalsize $^1$Chair of Structural Mechanics and Analysis, TU Berlin,} \\
    {\normalsize Straße des 17. Juni 135, 10623 Berlin, Germany} \\[1.5mm]
	{\normalsize $^2$Institute of Mechanics, TU Dortmund University,} \\
	{\normalsize Leonhard-Euler-Straße 5, 44227 Dortmund, Germany}\\ \\
	}

\date{{\small \today}}

\begin{document}

\maketitle

\begin{abstract}
The present contribution proposes a thermodynamically consistent model for the simulation of the ductile damage. The model couples the phase field method of fracture to the Armstrong-Frederick plasticity model with kinematic hardening. The latter is particularly suitable for simulating the material behavior under a cyclic load. The model relies on the minimum principle of the dissipation potential. However, the application of this approach is challenging since potentials of coupled methods are defined in different spaces: The dissipation potential of the phase field model is expressed in terms of rates of internal variables, whereas the Armstrong-Frederick model proposes a formulation depending on thermodynamic forces. For this reason, a unique formulation requires the Legendre transformation of one of the potentials. The present work performs the transformation of the Armstrong-Frederick potential, such that final formulation is only expressed in the space of rates of internal variables. With the assumption for the free energy and the joint dissipation potential at hand, the derivation of evolution equations is straightforward. The application of the model is illustrated by selected numerical examples studying the material response for different load constellations and sample geometries. The paper provides a comparison with the experimental results as well.
\end{abstract}


\section{Introduction}

Fatigue failure   significantly influences the safety measures during the service life of engineering structures. It particularly holds for structures subjected to alternate loading with high amplitudes, which is an often situation when speaking of transportation vehicles. Dependent on the number of cycles and load amplitude, the interplay of  the brittle and ductile failure takes place. 
%
%
However, the present contribution focuses on the low cycle fatigue where the role of plasticity is essential. 
%
The high complexity of the ductile damage under the cyclic load  strongly motivates the virtual testing of materials on the basis of the computer simulations. The present work envisages a coupling of the phase field method of fracture to the Armstrong-Frederick model of plasticity with the kinematic hardening to this end. 

The base for the phase field model of fracture has been set in the work by  \cite{ma40} proposing the formulation of brittle fracture based solely on Griffith's idea of competition between elastic and fracture energy. This work initiated a large group of further contributions related to the numerical implementation of the concept proposed  \citep{ma18,ma74, ma41, ma84}. Amongst others, Ambrosio-Tortorelli regularizations have become ubiquitous \citep{ma5, ma6, ma20}. These approaches are nowadays known as phase-field models of fracture and share several common features with the approaches resulting from Ginzburg-Landau models for phase transition \citep{ma57}.
These regularizations have  been applied to a wide variety of fracture problems including fracture of thermal and drying cracks \citep{ma68, ma22}, ferro-magnetic and piezo-electric materials \citep{ma1, ma90} and hydraulic fracturing \citep{ma19, ma89, ma91}. These models also been enhanced to account for cohesive effects \citep{ma33, ma42}, ductile behavior \citep{ma2, ma71, ma3}, large deformations \citep{ma4, ma70, ma14}, quasi-brittle damage \citep{ma92} and anisotropy \citep{ma63}.

Some of previous works have also considered coupling of plasticity to the phase field method \citep{ma2, ma3,ma4, ma14, ma70,ma71}. However, the current contribution will focus more precisely on the plasticity  under the cyclic load  where the Armstrong-Frederick model is a widely accepted concept \citep{r3,r15,r16,r25}. Some of the works in this field deal with so-called generalized representations, where either the classical Armstrong-Frederick model is extended to include several Armstrong-Frederick terms \citep{r3}, or where the application of multiple yield surfaces is envisaged \citep{r25}. Another frequently used variant of the Armstrong-Frederick model has been presented by \cite{r24} and by  \cite{r16}. The effects of the kinematic hardening are nowadays also investigated within the framework of large deformations \citep{r19,r38,r18,r23}.

The Armstrong-Frederick model coupled to the phase field method  is a promising approach inheriting the advantages of both incorporated techniques. However, the numerical implementation of this coupled approach brings with it  several challenges such as the definition of  a unique framework for both setups, the derivation of coupled evolution equations, the distinction between the tension and compression mode and  certainly  the development of the computationally efficient algorithm. Some of these issues are discussed in the present contribution which is structured as follows.  An overview on the phase field method of fracture is presented in Sect. 2, whereas Sect. 3 introduces the Armstrong-Frederick model of plasticity. Furthermore, the Armstrong-Frederick evolution equations are derived  by using two versions of the minimum principle of  dissipation potential (Sects. 4 and 5). The concept for the coupling of two methods is presented in Sect. 6 and complemented  by explaining the scheme for the calculation of  the plasticity multiplier on the basis of the consistency condition (Sect. 7). Details on the numerical implementation including the global part and the material point part are presented in Sect. 8.  Eventually, the paper also includes  numerical examples simulating the tests with the uniformly increasing   and  the cyclic load. The contribution finishes with conclusions and an outlook.

\section{Diffusive crack topology}\label{ph_field}
 The derivation of the coupled method requires a short recapitulation of separate approaches, which is done in two subsequent sections.  Sect. \ref{ph_field} provides an overview of the phase field method, whereas Sect. \ref{armstrongfrederick} focuses on the Armstrong-Frederick model.

The present work uses  the phase field  framework proposed by Miehe and coworkers \citep{miehe2010phase,miehe2010thermodynamically} as a basis. This approach supposes the diffusive (regularized)  crack topology instead of the  sharp crack topology leading to  serious difficulties in the numerical implementation due to the lack of the continuity and differentiability. It relies on the evaluation of the damage   variable $d$, which continuously changes in the range $[0,1]$, where zero-value corresponds to the intact material and the unity-value to the fully broken state. The approach starts by postulating an expression  for the crack surface density function depending on the damage variable  $d$

\begin{equation}\label{gammad}
	\gamma(d,\nabla d) = \frac{1}{2l}d^2 + \frac{l}{2}\lvert\nabla d\rvert^2,
\end{equation}
which is furthermore used to  define  the total crack surface $A$ for the entire body $\mathcal{B}$

\begin{equation}\label{endeq}
	A(d) = \int_\mathcal{B}\gamma(d,\nabla d) \, d V.
\end{equation}
An assumption that the crack propagation is a fully dissipative process yields to the conclusion that the constitutive dissipation potential has  to be proportional to the rate of crack surface density

\begin{equation} \label{dp}
	\Phi = g_\mathrm{c}\dot{\gamma}(\dot{d},\nabla\dot{d},d,\nabla d),
\end{equation}
where  parameter $g_\mathrm{c}$ is related to the critical Griffith-type fracture energy and can be seen as a constitutive threshold value.
 However, this potential still does not guarantee that the damage  is an increasing function.
 For this reason, the potential is extended  by introducing the penalty term
 
\begin{equation}\label{penalty}
	P(\dot{d}) = \frac{k_{\mathrm{p}}}{2}\langle\dot{d}\rangle^2_-.
\end{equation}
Here, the negative Macaulay brackets $\langle\bullet\rangle_- = (\bullet-\lvert\bullet\rvert)/2$ assure that the penalty term only activates for a negative damage evolution and  the constant $k_{\mathrm{p}}$ has to be chosen as high as possible in order to  stipulate the condition $\dot{d}\geq0$. 
Along with the penalty term \eqref{penalty}, the damage potential turns into

\begin{equation} \label{dp_pen}
	\Phi^{\mathrm {d}} = \Phi+ P(\dot{d})=g_\mathrm{c}\dot{\gamma}(\dot{d},\nabla\dot{d},d,\nabla d) + \frac{k_{\mathrm{p}}}{2}\langle\dot{d}\rangle^2_-.
\end{equation}
In a further step, the model focuses on the  definition of the free energy and its reduction on the basis of function $\omega=((1-d)^2 + k_{\mathrm{d}})$. Here, the positive  constant $k_{\mathrm{d}}$ prevents the energy to become identical to zero at a fully broken state. Moreover, the reduction  function only affects the  tension part of the energy which corresponds to the concept of the anisotropic degradation of energy

\begin{equation} \label{energysplit}
{\Psi}^{\mathrm{d}}(\boldsymbol{\epsilon},d) = \,\omega(d)\,{\Psi}_{+}(\boldsymbol{\epsilon}) + {\Psi}_{-}(\boldsymbol{\epsilon}).
\end{equation}
The additive split  of the energy  into a tension part ${\Psi}_{+}(\boldsymbol{\epsilon})$ and  a compression part ${\Psi}_{-}(\boldsymbol{\epsilon})$, is  of the special interest for simulating the cyclic behavior where the crack closure in the compression mode has to be considered. The definitions of the positive and negative energy parts of the free energy  are based on the spectral decomposition of the strain tensor

\begin{equation}\label{spectral}
	\boldsymbol{\epsilon}_\pm = \sum_i \langle\epsilon_i\rangle_\pm\boldsymbol{n}_i\otimes\boldsymbol{n}_i,
\end{equation}
where $\epsilon_i$ are the principal strains, $\boldsymbol{n}_i$ are the principal strain directions and $i$ is the summation index. Equation \eqref{spectral} uses  positive and negative Macaulay brackets. Positive ones are defined as $\langle\bullet\rangle_+ = (\bullet+\lvert\bullet\rvert)/2$, whereas the negative ones have already been used in Eq. \eqref{penalty}. The stresses corresponding to \eqref{spectral} are then defined as

\begin{equation} 
\boldsymbol{\sigma}=\frac{\partial \Psi^{\mathrm{d}}}{\partial{\boldsymbol{\epsilon}}}=\omega(d)\,\boldsymbol{\sigma}_{+ }+\boldsymbol{\sigma}_{-}
\end{equation}
such that it holds $\boldsymbol{\sigma}_{\pm}=\frac{\partial \Psi_{\pm}}{\partial\boldsymbol{\epsilon}}$.
For a case of an isotropic material characterized by Lam\'e constants $\lambda$ and $\mu$, the energy spit turns into 

\begin{equation}
	\Psi_\pm(\boldsymbol{\epsilon})=\frac{1}{2}\lambda(\mathrm{tr}\langle\boldsymbol{\epsilon}\rangle_{\pm})^2 + \mu\mathrm{tr}(\langle\boldsymbol{\epsilon}\rangle^2_\pm)
\end{equation}
with the corresponding equilibrium equation and stresses

\begin{gather}
\mathrm{Div}[\,\omega(d)\,\boldsymbol{\sigma}_{+}+\boldsymbol{\sigma}_{-}] = \boldsymbol{0}, \label{equil}
\\
\boldsymbol{\sigma}_\pm=\sum_{i}{[\lambda\mathrm{tr}\langle\boldsymbol{\epsilon}\rangle_\pm+2\mu\langle\epsilon_{i}\rangle_\pm]\, \boldsymbol {n}_{i}\otimes\boldsymbol{n}_{i}}.
\end{gather}
Since the free energy and the dissipation potential  are known, the minimum principle of dissipation potential is used to derive the  evolution equation for the internal parameter, namely

\begin{gather}
\underset{\dot{d}}{\text{min}}(\mathcal{L}^\mathrm{MDP} =\dot{\Psi}^{\mathrm{d}}(\dot{\boldsymbol{\epsilon}},\dot{d})+\Phi^{\mathrm{d}}(d,\dot{d})) \qquad\Rightarrow
\qquad
\frac{g_c}{l}[d-l^2\Delta d]+[\omega'\,\Psi_{+}+k_{\mathrm{p}}\langle\dot{d}\rangle_-]=0.
\label{ev_dem}
\end{gather}
The equilibrium equation \eqref{equil} together with the evolution equation \eqref{ev_dem}b defines the strong problem of brittle fracture.


\section{The Armstrong-Frederick kinematic hardening model} \label{armstrongfrederick}
The Armstrong-Frederick kinematic hardening model \citep{armstrong1966mathematical} is widely accepted to  simulate the  characteristic phenomena of the hardening behavior of metals, namely the Bauschinger and the ratcheting effect. Within the framework of the small strain plasticity, the formulation of the Armstrong-Frederick model starts with the typical assumption for the additive decomposition of strain tensor 
$\boldsymbol{\epsilon}$  into an elastic part $\boldsymbol{\epsilon}^\mathrm{e}$ and a plastic part $\boldsymbol{\epsilon}^\mathrm{p}$, namely $\boldsymbol{\epsilon} = \boldsymbol{\epsilon}^\mathrm{e} + \boldsymbol{\epsilon}^\mathrm{p}.$ However, the description of the hardening effects requires the introduction of an additional strain-like internal variable $\boldsymbol{\xi}$. 
According to Lion \citep{r18}, $\xi$ relates to local viscoelastic deformations induced by dislocations. In rheological models, it is simulated as a damping element connected in series to  an elastic spring. These are additionally coupled to a friction element in parallel \citep{dettmer2004theo}.
This internal variable is responsible for  the change of material stiffness depending on the loading history and is incorporated in  the Helmholtz free energy, which now includes two parts: the elastic energy $\Psi^{\mathrm{el}}(\boldsymbol{\epsilon}^{\mathrm{e}})$ and free energy due to the hardening $\Psi^{\mathrm{hard}}(\boldsymbol{\xi})$, both  in the quadratic form 

\begin{equation}
	\Psi^\mathrm{AF} =
\Psi^{\mathrm{el}}(\boldsymbol{\epsilon}^{\mathrm{e}})+	\Psi^{\mathrm{hard}}(\boldsymbol{\xi})=
	\frac{1}{2}(\boldsymbol{\epsilon}-\boldsymbol{\epsilon}^\mathrm{p}):\mathbb{C}:(\boldsymbol{\epsilon}-\boldsymbol{\epsilon}^\mathrm{p}) + \frac{1}{2}c \ \boldsymbol{\xi}:\boldsymbol{\xi}. \label{energyaf}
\end{equation}
Here, $\mathbb{C}$ is the fourth order elasticity tensor and $c$ represents the  kinematic hardening modulus. The Helmholtz free energy serves furthermore as a basis for the formulation of dissipation which, in the case of an isothermal process, only depends on the elastic power $(\boldsymbol{\sigma}:\dot{\boldsymbol\epsilon})$ and free energy rate $(\dot{\Psi}^\mathrm{AF})$

\begin{equation}
\mathcal{D} = \boldsymbol{\sigma}:\dot{\boldsymbol\epsilon}-\dot{\Psi}^\mathrm{AF} \geq 0. \end{equation}
In the concrete case of the Armstrong-Frederick model, the dissipation turns into:

\begin{equation}
(\boldsymbol{\sigma}-\frac{\partial\Psi^\mathrm{AF}}{\partial \boldsymbol{\epsilon}}):\dot{\boldsymbol{\epsilon}}{-\frac{\partial \Psi^\mathrm{AF}}{\partial \boldsymbol{\epsilon}^\mathrm{p}}}:\dot{\boldsymbol{\epsilon}}^\mathrm{p}{-\frac{\partial \Psi^\mathrm{AF}}{\partial \boldsymbol{\xi}}}:\dot{\boldsymbol{\xi}} \geq 0, 
\end{equation}
which yields  two groups of consequences. The first consequence is the constitutive law $\boldsymbol{\sigma}=\frac{\partial\Psi^\mathrm{AF}}{\partial \boldsymbol{\epsilon}}$, whereas the second consequence defines the driving forces of internal variables:

\begin{gather}
  \boldsymbol{q}_{\boldsymbol{\epsilon}^\mathrm{p}}= - \frac{\partial \Psi^\mathrm{AF}}{\partial \boldsymbol{\epsilon}^\mathrm{p}} = \mathbb{C}:[\boldsymbol{\epsilon}-\boldsymbol{\epsilon}^\mathrm{p}] = \boldsymbol{\sigma}, \\
  \boldsymbol{q}_{\boldsymbol{\xi}} = -\frac{\partial \Psi^\mathrm{AF}}{\partial\boldsymbol{\xi}}= -c \ \boldsymbol{\xi}=\boldsymbol{\chi} 
\label{chi}.
\end{gather}
The later definition corresponds to the  back stress and gives insight into the physical meaning of material parameter $c$. It is a proportionality constant relating the back stress $\chi$ to the strain-like quantity $\xi$. In analogy to the constitutive law of elastic materials, $c$ represents the kinematic hardening modulus.
A shorter notation for internal variables $\boldsymbol{\nu}=\{{\boldsymbol{\epsilon}^\mathrm{p}},{\boldsymbol{\xi}}\}$ and for corresponding driving forces $\boldsymbol{q}=\{\boldsymbol{\sigma}, \boldsymbol{\chi}\}$,  together with the constitutive law for stresses, yields the result for the so-called reduced dissipation 

\begin{equation}
 \mathcal{D}^\mathrm{red} = \boldsymbol{q}:\dot{\boldsymbol{\nu}}=\boldsymbol{\sigma}:\dot{\boldsymbol{\epsilon}}^\mathrm{p}+\boldsymbol{\chi}:\dot{\boldsymbol{\xi}}\geq 0. \label{dissipation}
\end{equation}
The Armstrong-Frederick model is eventually completed  by introducing the yield locus formula
 
\begin{equation} \label{yieldlocus}
\Omega^\mathrm{AF}  =\lvert \lvert \bsigma-\bchi \rvert \rvert 
	- \byield = 0 , \qquad \byield=\sqrt{2/3} \sigma_{\mathrm{Y}},
\end{equation}
 where symbol $\rvert\rvert \bullet\lvert\lvert = \sqrt{\bullet:\bullet}$ denotes the Frobenius norm, $\bsigma$ and $\bchi$ are deviatoric parts of stresses $\boldsymbol{\sigma}$ and back stresses $\boldsymbol{\chi}$ and $\sigma_\mathrm{Y}$ is the yield limit. Condition \eqref{yieldlocus} defines the admissible set of driving forces yielding inelastic deformations. 

\section{Derivation of the evolution equations based on the minimum principle for the dissipation potential in terms of driving forces} \label{AF_MDP}
The completion of the Armstrong-Frederick model previously described also requires  the derivation of evolution equations  which can be accomplished by using different concepts. One possibility is to follow the  minimum principle of dissipation potential (MDP)  as proposed in works by \cite{dettmer2004theo} and by \cite{AYGUN2020}

\begin{gather} 
	\underset{\boldsymbol{q}}{\text{min}}\left(\mathcal{L}^{\mathrm{MDP^*}}=
-\boldsymbol{q} : \dot{\boldsymbol{\nu}}
	+\Phi^\mathrm{AF^*}\left(\boldsymbol{\nu},\boldsymbol{q}\right)\right), \label{MDP} 
	\\
	\Phi^\mathrm{AF^*}(\boldsymbol{\nu},\boldsymbol{q})  = a \lvert \lvert \bsigma-\bchi \rvert \rvert  + \frac{1}{2b}\lvert \lvert\bchi \rvert \rvert^2. 
	\label{dispot}
\end{gather}
The dissipation potential \eqref{dispot} is inspired by the expression for the yield locus \eqref{yieldlocus} but additionally includes a term depending  on the norm of driving force $\boldsymbol{\chi}$. Symbols $a$ and $b$ denote material parameters. However, the subsequent derivations will show that parameter $a$ does not influence the evolution equations, whereas parameter $b$ plays an important role and represents pseudo-viscoelasticty. 
Superscript  $*$  indicates that  a formulation in terms of driving forces is chosen. 
The Lagrangian corresponding  to the minimization problem \eqref{MDP} and \eqref{dispot} has the form

\begin{eqnarray}
	\mathcal{L}^{\mathrm{MDP^*}} =
	-\boldsymbol{\sigma}:\dot{\boldsymbol{\epsilon}}^\mathrm{p}-\boldsymbol{\chi}:\dot{\boldsymbol{\xi}}+\Phi^\mathrm{AF^*}(\boldsymbol{\nu},\boldsymbol{q}),
\end{eqnarray}
which yields the following stationary conditions

\begin{eqnarray} \label{st_con}
\frac{\partial\mathcal{L}^{\mathrm{MDP^*}}}{\partial \boldsymbol{q}} = \left( \begin{array}{c} \frac{\partial\mathcal{L}^{\mathrm{MDP^*}}}{\partial\boldsymbol{\sigma}} \\[0.2cm] \frac{\partial\mathcal{L}^{\mathrm{MDP^*}}}{\partial\boldsymbol{\chi}} 
\\ \end{array}\right)=  
 \left( \begin{array}{c} -\dot{\boldsymbol{{\epsilon}}}^\mathrm{p} +  \frac{\partial\Phi^\mathrm{AF^*}}{\partial\boldsymbol{\sigma}} \\ 	-\dot{\boldsymbol{\xi}} + \frac{\partial\Phi^\mathrm{AF^*}}{\partial\boldsymbol{\chi}}  \\ \end{array}\right) = \boldsymbol{0}. 
\end{eqnarray}
The first equation in \eqref{st_con}  determines the evolution of plastic strains  and shows that the rate $\dot{\boldsymbol{{\epsilon}}}^\mathrm{p}$ is equal to the derivative of  dissipation potential with respect to corresponding driving force:
$
	\dot{\boldsymbol{{\epsilon}}}^\mathrm{p} = \frac{\partial\Phi^\mathrm{AF^*}}{\partial\boldsymbol{\sigma}}
	=\frac{\bsigma-\bchi}{\lvert\lvert\bsigma-\bchi\rvert\rvert}. 
$
However, the received derivative is not uniquely defined and only contains the information on the direction of the flow of the plastic strains, not on the magnitude. For that reason, the right-hand side expression is scaled by the plastic multiplier $\lambda$ which yields the standard solution

\begin{equation}\label{dot_epsp}
		\dot{\boldsymbol{{\epsilon}}}^\mathrm{p} = \frac{\partial\Phi^\mathrm{AF^*}}{\partial\boldsymbol{\sigma}}=\lambda \frac{\bsigma-\bchi}{\lvert\lvert\bsigma-\bchi\rvert\rvert} .
\end{equation}
The second equation in \eqref{st_con} determines the  evolution of the internal variable  $\boldsymbol{\xi}$. Here, the same argumentation as in Eq. \eqref{dot_epsp}  is used for the first term, whereas the second term is uniquely determined and does not need to be scaled with the multiplier $\lambda$

\begin{equation}\label{xidot}
\dot{\boldsymbol{\xi}} = \frac{\partial\Phi^\mathrm{AF^*}}{\partial\boldsymbol{\chi}} = -\lambda\frac{\bsigma-\bchi}{\lvert\lvert\bsigma-\bchi\rvert\rvert} +  \frac{1}{b}\bchi =-	\dot{\boldsymbol{{\epsilon}}}^\mathrm{p} + \frac{1}{b}\bchi 
\qquad\Rightarrow\qquad
\bchi=b(\dot{\boldsymbol{\xi}}+\dot{\boldsymbol{{\epsilon}}}^\mathrm{p} ).
\end{equation}
Equations \eqref{dot_epsp}  and  \eqref{xidot}a show the deviatoric character of internal variables ${\boldsymbol{{\epsilon}}}^\mathrm{p}$ and ${\boldsymbol{\xi}}$. Moreover, Eq. \eqref{xidot}b represents a constitutive law typical of a viscous material with the viscosity $b$.    Finally, Eq. \eqref{chi} along with Eq. \eqref{xidot}a provides the evolution equation  for back stress $\chi$ which can also  be identified as a deviatoric quantity:

\begin{equation}
\dot{\boldsymbol {\chi}}=-c\dot{\boldsymbol{\xi}}=c\dot{\boldsymbol{\epsilon}}^\mathrm{p} - \frac{c}{b}\bchi=\dot{\bar{\boldsymbol {\chi}}}.
\end{equation}
The MDP-approach differs from the common approach  treating the yield locus equation $\Omega$ as a subsidiary condition  within the Lagrange formalism of the constrained optimization. The main difference manifests itself in the evolution equations taking the  form $\nobreak{ \dot{\boldsymbol{\nu}} = \lambda\frac{\partial\Omega}{\partial\boldsymbol{q}}}$ with $\lambda$ as the Lagrange multiplier. An important advantage of the MDP-approach compared to the constrained optimization method is that it allows a standardized variable transformation, which is often a useful tool in coupling strategies.

\section{Derivation of the evolution equations based on the minimum principle for the dissipation potential in terms of rates of internal variables}\label{sect_lengendre}

Both models summarized in Sects. \ref{ph_field} and \ref{AF_MDP} apply the minimum principle of dissipation potential. However, the phase field model uses a formulation in terms of the rate of internal variables,  whereas the  Armstrong-Frederick model minimizes the dissipation potential in terms of driving forces. Naturally, the coupling procedure requires a unique formulation in a single space. The present contribution deals with the approach in terms of velocities, such that the transformation of the potential $\Phi^{\mathrm{AF^*}}(\boldsymbol{\nu}, \boldsymbol{q})$ into the space of rates is necessary. This type of exchange is conducted on the basis of Legendre transformation (LT) defining the new potential as follows 

\begin{gather}
	\Phi^{\mathrm{AF}}(\boldsymbol{\nu},\dot{\boldsymbol{\nu}})=\underset{{\boldsymbol{q}}}{\text{max}}\{\mathcal{L}^\mathrm{LT}=\boldsymbol{q}:\dot{\boldsymbol{\nu}}-\Phi^{\mathrm{AF^*}}(\boldsymbol{\nu},{\boldsymbol{q}})\},\label{legendre} 
	\\
	\mathcal{L}^\mathrm{LT} = \boldsymbol{\sigma}:\dot{\boldsymbol{{\epsilon}}}^\mathrm{p}+\boldsymbol{\chi}:\dot{\boldsymbol{\xi}} - a||\bsigma-\bchi||-\frac{1}{2b}||\bchi||^2. 
	\label{llt}
\end{gather}
The maximization procedure relies on two stationary conditions, the first of which gives a relationship for rates of plastic deformations

\begin{equation}
	\frac{\partial \mathcal{L}^\mathrm{LT} }{\partial\boldsymbol{\sigma}}=\dot{\boldsymbol{{\epsilon}}}^\mathrm{p}-\lambda\frac{\bsigma-\bchi}{\lvert\lvert\bsigma-\bchi\rvert\rvert}= 0 . \label{stat1}
\end{equation}
Bearing in mind that the multiplier $\lambda$ is a scalar, Eq. \eqref{stat1} shows that rate $\dot{\boldsymbol{{\epsilon}}}^\mathrm{p}$ and difference $\bsigma-\bchi$ are coaxial. In other words, it holds that
$
	\frac{\dot{\boldsymbol{{\epsilon}}}^\mathrm{p}}{||\dot{\boldsymbol{{\epsilon}}}^\mathrm{p}||}=\frac{\bsigma-\bchi}{\lvert\lvert\bsigma-\bchi\rvert\rvert}.
$
Moreover, the plastic flow only occurs if the stress state fulfills the yield locus equation \eqref{yieldlocus} such that the norm $	||\bsigma-\bchi||$ can be replaced by the yield limit $\byield$.
Equation \eqref{stat1}	is then rewritten as follows

	\begin{equation}
\bsigma-\bchi=\byield\frac{\dot{\boldsymbol{{\epsilon}}}^\mathrm{p}}{||\dot{\boldsymbol{{\epsilon}}}^\mathrm{p}||}
\qquad \Rightarrow \qquad
\bsigma=\byield\frac{\dot{\boldsymbol{{\epsilon}}}^\mathrm{p}}{||\dot{\boldsymbol{{\epsilon}}}^\mathrm{p}||}+\bchi. \label{sig}
\end{equation}
On the other hand, the second stationary condition 

\begin{equation}
	\frac{\partial  \mathcal{L}^\mathrm{LT}}{\partial \boldsymbol{\chi}}=\dot{\boldsymbol{\xi}}+\lambda\frac{\bsigma-\bchi}{||\bsigma-\bchi||}-\frac{1}{b}\bchi=0 
	\qquad \Rightarrow \qquad
\bchi=b\dot{\boldsymbol{\xi}}+b\dot{\boldsymbol{{\epsilon}}}^\mathrm{p} \label{chii}
	\end{equation}
together  with \eqref{sig}b leads to the final expression for deviatoric stresses

\begin{equation}
	\bsigma=\byield\frac{\dot{\boldsymbol{{\epsilon}}}^\mathrm{p}}{||\dot{\boldsymbol{{\epsilon}}}^\mathrm{p}||}+b\dot{\boldsymbol{\xi}}+b\dot{\boldsymbol{{\epsilon}}}^\mathrm{p} \label{newsig}.
\end{equation}
Driving forces \eqref{chii}b and \eqref{newsig} are now  inserted into \eqref{llt}, which reads the desired  dissipation potential in terms of the internal variables and their rates 

\begin{equation}
\Phi^{\mathrm{AF}}(\boldsymbol{\nu},\dot{\boldsymbol{\nu}})=\byield||\dot{\boldsymbol{{\epsilon}}}^\mathrm{p}||
+\frac{b}{2}||\dot{\boldsymbol{{\epsilon}}}^\mathrm{p}||^2+\frac{b}{2}||\dot{\boldsymbol{\xi}}||^2+b\,\dot{\boldsymbol{{\epsilon}}}^\mathrm{p}\!:\!\dot{\boldsymbol{\xi}}.  \label{klingepot}
\end{equation}
More details on derivation are presented in Appendix \ref{appendix_a}.
Result \eqref{klingepot} now enables the formulation of the new minimization problem

\begin{equation}
\underset{\dot{\boldsymbol{\nu}}}{\text{min}}\left(\mathcal{L}^\mathrm{MDP} =\dot{\Psi}^{\mathrm{AF}}+\Phi^{\mathrm{AF}}(\boldsymbol{\nu},\dot{\boldsymbol{\nu}})\right), \label{MDP_nu}
\qquad
		\mathcal{L}^\mathrm{MDP}=\boldsymbol{\sigma}:\dot{\boldsymbol{\epsilon}}-\boldsymbol{\sigma}:\dot{\boldsymbol{\epsilon}}^\mathrm{p}-\boldsymbol{\chi}:\dot{\boldsymbol{\xi}}+\Phi^{\mathrm{AF}}(\boldsymbol{\nu},\dot{\boldsymbol{\nu}}). 
\end{equation}
Bearing in mind the deviatoric character of quantities $\dot{\boldsymbol{\epsilon}}$ and $\dot{\boldsymbol{\xi}}$, the problem can also be written as:

 \begin{equation}
\underset{\dot{\boldsymbol{\nu}}}{\text{min}}\left(\mathcal{L}^\mathrm{MDP} =\dot{\Psi}^{\mathrm{AF}}+\Phi^{\mathrm{AF}}(\boldsymbol{\nu},\dot{\boldsymbol{\nu}})\right), \label{MDP_nu1}
\qquad
		\mathcal{L}^\mathrm{MDP}=\boldsymbol{\sigma}:\dot{\boldsymbol{\epsilon}}-\bsigma:\dot{\boldsymbol{\epsilon}}^\mathrm{p}-\bchi:\dot{\boldsymbol{\xi}}+\Phi^{\mathrm{AF}}(\boldsymbol{\nu},\dot{\boldsymbol{\nu}}). 
\end{equation}

Equation  \eqref{MDP_nu1} is the counterpart of  the MDP-problem \eqref{MDP} in terms of velocities and provides  expressions  for driving forces  according to relationship $\boldsymbol {q}=\frac{\partial \Phi^{\mathrm{AF}}}{\partial{\dot{\boldsymbol{\nu}}}}$:

\begin{gather}
\boldsymbol{q}_{\boldsymbol{\epsilon}^{\mathrm{p}}}=\bsigma=\frac{\partial \Phi^{\mathrm{AF}}}{\partial{\dot{\boldsymbol{\epsilon}}^\mathrm{p}}}=\byield\frac{\dot{\boldsymbol{{\epsilon}}}^\mathrm{p}}{||\dot{\boldsymbol{{\epsilon}}}^\mathrm{p}||}+ b{\dot{\boldsymbol{{\epsilon}}}^\mathrm{p}}+b{\dot{\boldsymbol{{\xi}}}}, \label{q_epsp}
\\
\boldsymbol{q}_{\boldsymbol{\xi}}=\bchi= \frac{\partial \Phi^{\mathrm{AF}}}{\partial{\dot{\boldsymbol{\xi}}}}= b{\dot{\boldsymbol{{\xi}}}}+b{\dot{\boldsymbol{{\epsilon}}}^\mathrm{p}} . \label{q_xi}
\end{gather}
A transformation of the system \eqref{q_epsp} and \eqref{q_xi} yields the evolution equations identical to the ones from Sect. \ref{AF_MDP} (Eqs. \eqref{dot_epsp} and \eqref{xidot}):

\begin{equation}\label{ev_eq_one}
{\dot{\boldsymbol{{\epsilon}}}^\mathrm{p}}=\lambda \frac{\bsigma-\bchi}{||\bsigma-\bchi||}, 
\qquad 
\dot{\boldsymbol{\xi}}=-{\dot{\boldsymbol{{\epsilon}}}^\mathrm{p}}+\frac{1}{b} \bchi.
\end{equation}
Note that  the insertion of \eqref{q_xi} into \eqref{q_epsp} yields the  intermediate result

\begin{equation}
\bsigma=\byield\frac{\dot{\boldsymbol{{\epsilon}}}^\mathrm{p}}{||\dot{\boldsymbol{{\epsilon}}}^\mathrm{p}||} + \boldsymbol{\chi}
\qquad\Rightarrow\qquad 
\bsigma-\bchi=\byield\frac{\dot{\boldsymbol{{\epsilon}}}^\mathrm{p}}{||\dot{\boldsymbol{{\epsilon}}}^\mathrm{p}||}, \label{cons_two}
\end{equation}
which is only valid  for $||\dot{\boldsymbol{{\epsilon}}}^\mathrm{p}||\neq 0$. By taking the norm of \eqref{cons_two}b, it follows that $ {||\bsigma-\bchi||}=\byield$. This consequence indicates that dissipation potential \eqref{klingepot} intrinsically includes the yield locus condition. 
Accordingly, it can be summed up that the dissipation potential \eqref{klingepot} describes the same problem as the dissipation potential \eqref{dispot} along with the yield locus function \eqref{yieldlocus}.

Finally, the strong form corresponding to the Armstrong-Frederick problem can be recapitulated as follows: 

\begin{eqnarray}
&&\mathrm{Div}\,\boldsymbol{\sigma}=\boldsymbol{0} ,\\
&&\boldsymbol{\sigma} = \mathbb{C}:(\boldsymbol{\epsilon}-\boldsymbol{\epsilon}^\mathrm{p}), \label{sigmaeq} \\
&&\dot{\boldsymbol{\epsilon}}^\mathrm{p}  = \lambda \frac{\bsigma-\bchi}{||\bsigma-\bchi||}, \label{evolutionepsp} \\
&&\dot{\boldsymbol{\chi}}= c\left(\dot{\boldsymbol{\epsilon}}^\mathrm{p}- \ \frac{1}{b} \bchi\right ) \label{chidot}, \\
&&\Omega^{\mathrm{AF}} = ||\bsigma- \bchi|| -\byield \label{phi}, \\
&&\lambda \geq 0, \ \Omega^{\mathrm{AF}} \leq 0, \ \lambda\Omega^{\mathrm{AF}} = 0. \label{kkk}
\end{eqnarray} 
The previous system includes the equilibrium equation, the constitutive law, two evolution equations and the Karush-Kuhn-Tucker conditions defining the plastic domain. Equation \eqref{chidot} is derived by using the time derivative of definition \eqref{chi} and rate \eqref{ev_eq_one}b. The body forces in the equilibrium equation are neglected.  Evolution equation \eqref{evolutionepsp} and Karush-Kuhn-tucker  conditions depend on plastic multiplier $\lambda$ which is typically determined from the consistency condition.

\section{Coupling of the Armstrong-Frederick model to the phase-field approach}\label{coupl_sec}
The coupling of the two methods starts by writing the Armstrong-Frederick energy in a form splitting the tension and compression part  of elastic energy in order to introduce  the damage influence  as was done in Sect. \ref{ph_field}

\begin{equation}
\Psi^{\mathrm{AF}}(\boldsymbol{\epsilon}^\mathrm{e},\boldsymbol{\xi})=\Psi^{\mathrm{el}}(\boldsymbol{\epsilon}^\mathrm{e})+\Psi^{\mathrm{hard}}(\boldsymbol{\xi})=\Psi^{\mathrm{el}}_{+}(\boldsymbol{\epsilon}_{}^\mathrm{e})+\Psi^{\mathrm{el}}_{-}(\boldsymbol{\epsilon}_{}^\mathrm{e})+\Psi^{\mathrm{hard}}(\boldsymbol{\xi}).
\end{equation}
However, damage also influences the hardening energy in the case where the tension mode is active. For this reason, the present model introduces a function distinguishing the pure compression mode from the pure tension and from the mixed modes 

\begin{equation}
  \bar{\omega} = 
  \begin{cases}
    \omega(d) & \text{if} \quad \text{max}\{(\epsilon_i)_{i=1,2,3},0\}>0 ,\\
    1 & \text{if} \quad \text{max}\{(\epsilon_i)_{i=1,2,3},0\}\leq 0 .
  \end{cases}
  \label{omegabar}
\end{equation}
By using this new notation, the coupled free energy is written as

\begin{equation}\label{coupen}
\Psi^{\mathrm{c}}(\boldsymbol{\epsilon}^\mathrm{e},d,\boldsymbol{\xi})=\omega\Psi^{\mathrm{el}}_{+}(\boldsymbol{\epsilon}_{}^\mathrm{e})+\Psi^{\mathrm{el}}_{-}(\boldsymbol{\epsilon}_{}^\mathrm{e})+\bar{\omega}\Psi^{\mathrm{hard}}(\boldsymbol{\xi})
\end {equation}
and corresponding constitutive laws  take the form

\begin{gather}
\boldsymbol{\sigma}=\frac{\partial\Psi^{\mathrm{c}}}{\partial\boldsymbol{\epsilon}}=\omega\frac{\partial\Psi^{\mathrm{el}}_{+}}{\partial\boldsymbol{\epsilon}}
+\frac{\partial\Psi^{\mathrm{el}}_{-}}{\partial\boldsymbol{\epsilon}}=\omega\boldsymbol{\sigma}_{+}+\boldsymbol{\sigma}_{-},
\label{stresscoup}\\
\boldsymbol{\sigma}_\pm=\frac{\partial\Psi^{\mathrm{el}}_{\pm}}{\boldsymbol{\epsilon}}=
\mathbb{C}:\boldsymbol{\epsilon}_{\pm}^{\mathrm{e}}=
\mathbb{C}:\langle\boldsymbol{\epsilon}-\boldsymbol{\epsilon}^{\mathrm{p}}\rangle_{\pm},\\
\boldsymbol{\chi}=\frac{\partial\Psi^{\mathrm{c}}}{\partial{\boldsymbol{\xi}}}=-\bar{\omega} c \boldsymbol{\xi},
\label{backstr}
\end{gather}
where the spectral decomposition of elastic strains has the standard form

\begin{equation}
\boldsymbol{\epsilon}_{\pm}^{\mathrm{e}}=(\boldsymbol{\epsilon}-\boldsymbol{\epsilon}^{\mathrm{p}})_{\pm}=\sum_i\langle\epsilon^{\mathrm{e}}_{i}\rangle_{\pm} \boldsymbol{n}_i\otimes\boldsymbol{n}_i=\sum_i\langle(\boldsymbol{\epsilon}-\boldsymbol{\epsilon}^{\mathrm{p}})_{i}\rangle_{\pm} \boldsymbol{n}_i\otimes\boldsymbol{n}_i.
\end {equation}
Bearing in mind stress definition  \eqref{stresscoup}, the equilibrium equation turns into

\begin{equation}
\mathrm{Div}[\omega(d)\boldsymbol{\sigma}_{+}+\boldsymbol{\sigma}_{-}] = \boldsymbol{0}. \label{damageone}
\end{equation}
The damage influence on the dissipation functional related to the plastic deformations follows the same argumentation as in the case of the  hardening energy, such that the coupled dissipation potential $\Phi^{\mathrm{c}}$ consists of two terms where the second term is weighted by the function $\bar{\omega}$

\begin{equation}\label{coup_diss_pot}
\Phi^{\mathrm{c}}=\Phi^{\mathrm{d}}+\bar{\omega}\,\Phi^{\mathrm{AF}}.
\end{equation}
With the definitions \eqref{coupen} and \eqref{coup_diss_pot}, the minimization of the corresponding Lagrange  function

\begin{equation} \label{LFC}
\mathcal{L}^{\mathrm{MDP}}=\dot{\Psi}^{\mathrm{c}}+\Phi^{\mathrm{c}}=
\dot{\Psi}^{\mathrm{c}}+\Phi^{\mathrm{d}}+\bar{\omega}\Phi^{\mathrm{AF}}
\end{equation}
yields  the following system of equations which, together with equilibrium equation \eqref{damageone}, defines the strong form of the coupled problem

\begin{eqnarray}
&&	\boldsymbol{\sigma}= \omega(d)\boldsymbol{\sigma}_{+}+\boldsymbol{\sigma}_{-}, \\
&&\dot{\boldsymbol{\epsilon}}^\mathrm{p}  = \lambda \bN,\qquad \bN=\frac{\bsigma-\bchi}{||\bsigma-\bchi||}, \label{ev_pl} \\
&&\dot{\boldsymbol{\xi}}=\frac{1}{b\bar{\omega}}\bchi-\dot{\boldsymbol{\epsilon}}^\mathrm{p} ,\label{ev_xi}\\
&&\dot{\boldsymbol{\chi}}=-\dot{\bar{\omega}}c\boldsymbol{\xi}+ c\bar{\omega}\left(\dot{\boldsymbol{\epsilon}}^\mathrm{p}-\  \frac{1}{b\bar{\omega}} \  \bchi\right ),  \label{ev_chi}\\
&&\Omega^\mathrm{c} = ||\bsigma - \bchi|| -\bar{\omega}\byield,  \\
&&\lambda \geq 0, \ \Omega^\mathrm{c} \leq 0, \ \lambda\Omega^\mathrm{c} = 0, \label{kkkc}\\
&& \frac{g_c}{l}[d-l^2\Delta d]+[\omega'\,\Psi^\mathrm{el}_{+}+\bar{\omega}'\,\Psi^{\mathrm{hard}} +k_{\mathrm{p}}\langle\dot{d}\rangle_-]=0. \label{damagetwo}
\end{eqnarray} 
Here, a shorter notation $\bN=\frac{\bsigma-\bchi}{||\bsigma-\bchi||}$ is introduced to denote  the so-called direction tensor. Details on the derivation of  evolution equations and of the yield criterion are provided in Appendix \ref{appendix_b}.

\section{Determination of the plastic multiplier}\label{pl_multi}
Evolution equation \eqref{ev_pl} and the Karush-Kuhn-Tucker conditions \eqref{kkkc} depend on plastic multiplier  $\lambda$ which is commonly determined from the  consistency condition $\lambda\dot{\Omega}^\mathrm{c}=0$. An appropriate form of this condition is obtained  by taking the time derivative of the yield locus formula 

\begin{equation}\label{cons_cond}
	\lvert\lvert\bsigma-\bchi\rvert\rvert^2 = \bar{\omega}^2\byield^2
 \qquad \Rightarrow \qquad
       	(\bsigma-\bchi):(\dot{\bar{\boldsymbol{\sigma}}}-\dot{\bar{\boldsymbol{\chi}}}) = 0.
\end{equation}
Since difference $(\bsigma-\bchi)$ is deviatoric, the condition above also can be written as follows:

\begin{equation}\label{cons_cond1}
(\bsigma-\bchi):(\dot{\boldsymbol{\sigma}}-\dot{\boldsymbol{\chi}}) = 0.
\end{equation}
In a further step, the stress rate

\begin{equation}
\dot{\boldsymbol{\sigma}}=\omega'\,\boldsymbol{\sigma}_{+}+ \omega (d) \,\mathbb{C}\!:\!(\dot{\boldsymbol{\epsilon}}-\dot{\boldsymbol{\epsilon}}^{\mathrm{p}})_{+}+ \mathbb{C}\!:\!(\dot{\boldsymbol{\epsilon}}-\dot{\boldsymbol{\epsilon}}^{\mathrm{p}})_{-}
\end{equation}
along  with Eqs.  \eqref{ev_pl}  and \eqref{ev_chi} is introduced in  \eqref{cons_cond1} which yields

\begin{equation}
(\bsigma-\bchi):\left[\omega'\,\boldsymbol{\sigma}_{+}+\omega\,\mathbb{C}\!:\![\dot{\boldsymbol{\epsilon}}-\lambda\bN]_{+}+ \mathbb{C}\!:\![\dot{\boldsymbol{\epsilon}}-\lambda\bN]_{-}  
+\dot{\bar{\omega}}\,c\boldsymbol{\xi}
- c\bar{\omega}\lambda\bN+\frac {c}{b}\bchi\right]=0 .
\end{equation}
However, the split of the elastic strain rate into a positive and a negative part cannot be performed as a superposition of positive/negative parts of the total and plastic strains. For this reason, the previous equation can only be solved numerically by using an iterative procedure. For this purpose, the initial guess $\lambda_0$ is obtained by neglecting the terms including the stress rates

\begin{equation}\label{simp_prob}
(\bsigma-\bchi):\left[\omega'\,\boldsymbol{\sigma}_{+} 
+\dot{\bar{\omega}}\,c\boldsymbol{\xi}
- c\bar{\omega}\lambda_0\bN+\frac {c}{b}\bchi\right]=0.
\end{equation}
Thereafter, all subsequent steps follow the iteration rule for the calculation of the updated value $\lambda_i$

\begin{equation}
(\bsigma-\bchi):\left[\omega'\,\boldsymbol{\sigma}_{+}+\omega\,\mathbb{C}\!:\![\dot{\boldsymbol{\epsilon}}-\lambda_{i-1}\bN]_{+}+ \mathbb{C}\!:\![\dot{\boldsymbol{\epsilon}}-\lambda_{i-1}\bN]_{-}
+\dot{\bar{\omega}}\,c\boldsymbol{\xi}
- c\bar{\omega}\lambda_i\bN+\frac {c}{b}\bchi\right]=0.
\end{equation}
The iterative process stops if the prescribed accuracy is  achieved.

\section{Numerical implementation}
\subsection{General approach}
The minimum principle of dissipation potential along with the time-incremental variational principles represents an important tool for the numerical solution  of boundary value problems. Within this concept, the Lagrange function $\cal{L}^{\mathrm {MDP}}$ corresponding to the body $\Omega$ is integrated over a single time increment $[t_n,t_{n+1}]$. In its original form, this integral depends on external variables ($\boldsymbol{u}$), internal variables $(\boldsymbol{\nu})$  and on the rates of internal variables, so-called velocities $(\dot{\boldsymbol{\nu}})$. However, the velocities can be replaced  by the  Euler-forward approximation $\dot{\boldsymbol{\nu}}=(\boldsymbol{\nu}_{n+1}-\boldsymbol{\nu}_{n})/\Delta t$, where $\Delta t= t_{n+1}-t_{n}$ is the time increment. As a consequence, the result only depends on discrete values of  external and internal variables such that the time integration is performed as follows

\begin{equation} \label{discr}
\int_{t_n}^{t_{n+1}}\!\!\int_{\Omega} {\mathcal{L}^\mathrm{MDP}} dV dt\approx
\int_{\Omega}\left\{
\Psi(\boldsymbol{\epsilon}^{\mathrm{e}}_{n+1},\boldsymbol{\nu}_{n+1})-
\Psi(\boldsymbol{\epsilon}^{\mathrm{e}}_{n},\boldsymbol{\nu}_{n})+
 \Delta t\, \Phi(\boldsymbol{\nu}_{n+1},(\boldsymbol{\nu}_{n+1}-\boldsymbol{\nu}_{n+1})/\Delta t)
\right\} dV.
\end{equation}
Here, displacement $\boldsymbol{u}$ is  the only external variable since an isothermal process is considered. In a further step, the minimization with respect to the displacements and internal variables in the current time step $n+1$ yields the sought solution. Within this procedure, term $ \Psi(\boldsymbol{\epsilon}^{\mathrm{e}}_{n},\boldsymbol{\nu}_{n})$ can be neglected since it only depends on values in the previous time step $n$. By adding the  potential of external forces $l(t_{n+1},\boldsymbol{u}_{n+1})$ in \eqref{discr}, the new,  combined  Lagrangian is constructed

\begin{equation} \label{l_comb}
\mathcal{L}^{\mathrm{comb}}=
\int_{\Omega}
\left\{
\Psi(\boldsymbol{\epsilon}^{\mathrm{e}}_{n+1},\boldsymbol{\nu}_{n+1})
+
 \Delta t\, \Phi(\boldsymbol{\nu}_{n+1},(\boldsymbol{\nu}_{n+1}-\boldsymbol{\nu}_{n+1})/\Delta t)
\right\} dV
+l(t_{n+1},\boldsymbol{u}_{n+1}).
\end{equation}
Within the present work, the minimization of \eqref{l_comb} is performed by an approach consisting of two parts: the global level solution part and the material point solution part. The former calculates   deformation and  damage, whereas the latter evaluates the internal variables $\boldsymbol{\epsilon}_{\mathrm{p}}$ and $\boldsymbol{\xi}$ by using the predictor-corrector scheme. In the continuation, each solution part is explained separately. The indexes related to step $n+1$ are omitted in order to achieve a concise representation.


\subsection{Global solution part}
The definition of the global solution part (gl) starts  with the reduced formulation of the Lagrangian by only including the reduced elastic energy and damage potential

\begin{equation}
\mathcal{L}^{\mathrm{gl}}=\int_{\Omega}\mathcal{L} dV=
\int_{\Omega}\omega(d_{})\Psi_{+}^{\mathrm{el}}(\boldsymbol{\epsilon}^{\mathrm{e}}_{})
+\Psi_{-}^{\mathrm{el}}(\boldsymbol{\epsilon}^{\mathrm{e}}_{})dV
+\int_{\Omega}\Delta t\,
\Phi^{\mathrm{d}}(d_n,\nabla d_n, d_{},\nabla d_{}) dV
+l(t_{},\boldsymbol{u}_{}).
\end{equation}
By using the expression for the crack surface density \eqref{gammad} and definition \eqref{dp_pen} the increment $\Delta \Phi^{\mathrm{d}}$ can  be reconstructed as shown in \cite{miehe2010phase,miehe2010thermodynamically}

\begin{equation}
\Delta t\, \Phi^{\mathrm{d}}(d_n,\nabla d_n, d_{},\nabla d_{})=\frac{g_c}{2l}(d^2-d_n^2)+\frac{g_c l}{2}(|\nabla d|^2-|\nabla d_{n}|^2)+\frac{k_{\mathrm{p}}}{2\Delta t}\langle d-d_n\rangle^2_{-}.
\end{equation}
The FE-implementation now requires the derivatives of the Lagrangian $\mathcal{L}$. The first derivatives are required in order to form the residual

\begin{gather} 
\partial_{\boldsymbol{\epsilon}} \mathcal{L} = \omega\,\mathbb{C}\!:\!\boldsymbol{\epsilon}^{\mathrm{e}}_{+}+
\mathbb{C}\!:\!\boldsymbol{\epsilon}^{\mathrm{e}}_{-},
\label{reps}\\
\partial_{d} \mathcal{L}=\omega'\,\,\Psi^{\mathrm{el}}_{+} +\frac{g_c}{l}d+\frac{k_{\mathrm{p}}}{\Delta t}\langle  d-d_n \rangle_{-}, \qquad  \partial_{\nabla d} \mathcal{L}=g_c l\nabla  d,
\end{gather}
whereas the second derivatives are needed for the definition of the stiffness matrix

\begin{gather}
\partial^2_{\boldsymbol{\epsilon}\boldsymbol{\epsilon}} \mathcal{L}=\omega \,\mathbb{C}\!:\!\boldsymbol{I}_{+}^{\boldsymbol{\epsilon}^\mathrm{e}}+ \mathbb{C}\!:\!\boldsymbol{I}_{-}^{\boldsymbol{\epsilon}^\mathrm{e}}, \qquad
\hspace{12.5mm}
\partial^2_{\boldsymbol{\epsilon}d}\mathcal {L}=\omega'\,\mathbb{C}\!:\!\boldsymbol{\epsilon}_{+}^{\mathrm{e}},
\\
\partial^2 _{dd}\mathcal {L}= \omega''\,\Psi_{+}^{\mathrm{el}}+ \frac{g_c}{l} +\frac{k_{\mathrm{p}}}{\Delta t}I^{d},\qquad \qquad \hspace{2mm}
\partial^2_{\nabla d \nabla d}\mathcal{L}=g_c l.
\end{gather}
These relationships use the following indicator functions

\begin{gather}
\boldsymbol{I}_{+}^{\boldsymbol{\epsilon}^\mathrm{e}}=\frac{\partial\boldsymbol{\epsilon}^\mathrm{e}_{+}}{\partial\boldsymbol{\epsilon}^\mathrm{e}}=\sum_{i} I_{i+}^{\mathrm{e}}\boldsymbol{m}_i\otimes \boldsymbol{m}_i,
\qquad \boldsymbol{m}_i=\boldsymbol{n}_i\otimes\boldsymbol{n}_i,
\\
\boldsymbol{I}_{-}^{\boldsymbol{\epsilon}^\mathrm{e}}=\frac{\partial\boldsymbol{\epsilon}^\mathrm{e}_{-}}{\partial\boldsymbol{\epsilon}^\mathrm{e}}=\sum_{i} (1-I_{i+}^{\mathrm{e}})\boldsymbol{m}_i\otimes \boldsymbol{m}_i,\\
I_{i+}^{\mathrm{e}}=
\begin{cases}
    1 & \text{if} \quad\epsilon_i>0 , \\
    0 & \text{if} \quad\epsilon_i\leq 0,
  \end{cases}
	\qquad
I^{d}=
\begin{cases}
    1 & \text{if} \quad d<d_n ,     \\
    0 & \text{if} \quad d \geq d_n .
  \end{cases}
	\end{gather}
The remaining part of the numerical procedure at this level encompasses the standard steps typical of  an FE-model. To this end, the bilinear shape functions  corresponding to a quadrilateral element and staggered solution scheme are applied in the present work. This approach staggers between the displacement and the phase-field and is advantageous compared to the  monolithic  solution type due to its higher robustness. However, both approaches are well established nowadays as shown in  works applying  monolithic schemes  \citep{miehe2010thermodynamically,Kuhn10,Schlueter14,Msekh15} or using its counterpart 
\citep{miehe2010phase,Borden12, Hofacker13}. The standard Newton-Raphson procedure is applied for the solution of nonlinear systems of equations.


\subsection{Material point solution part}
The material point solution part evaluates internal variables $\boldsymbol{\epsilon}^{\mathrm{p}}$ and $\boldsymbol{\xi}$ by using  a predictor-corrector scheme if damage and deformations $(d,\boldsymbol{u})$ as well as rate $\dot{\bar{\omega}}= \omega'\, \frac{(d-d_n)}{\Delta t}$ are known from the global level  solution. 
The predictor step calculates the trial stress  by assuming that  plastic deformations do not change in comparison with the previous time step $n$ 

\begin{equation}\label{predict}
	\boldsymbol{\sigma}^\mathrm{tr}=\omega(d) \ \mathbb{C}\!:\!(\boldsymbol{\epsilon}_{}-\boldsymbol{\epsilon}^\mathrm{p}_{n})_{+} +\mathbb{C}\!:\!(\boldsymbol{\epsilon}_{}-\boldsymbol{\epsilon}^\mathrm{p}_{n})_{-}.
\end{equation}
The trial stress is furthermore introduced in the yield locus function to check whether the prediction is true 

\begin{equation}
	\Phi^{\mathrm{c,tr}} = ||	\bsigma^\mathrm{tr} - \bchi_{n}|| - \bar{\omega} \ \byield.
\end{equation}
If the control value $\Phi^{\mathrm{c,tr}}$ is negative, the prediction \eqref{predict} is correct and the plastic deformations do not evolve in the present step. Otherwise,   a corrector step is needed to update the plastic deformations. The corrector step firstly calculates the plastic multiplier according to the scheme shown in Sect. \ref{pl_multi}.
The initial value $\lambda_0$ is determined by solving the simplified problem \eqref{simp_prob}

\begin{gather}
\lambda_0=\frac{(\bsigma^{\mathrm{tr}}-\bchi_n):\left[\omega'\,\bsigma_+^{\mathrm{tr}}
+\dot{\bar{\omega}}\,c\boldsymbol{\xi}_n
+\frac {c}{b}\bchi_n\right]}
{c \bar{\omega}\,(\bsigma^{\mathrm{tr}}-\bchi_n):\bN^{\mathrm{tr}}},
\end{gather}
whereas the later iterations follow the rule

\begin{gather}
\lambda_i=\frac{(\bsigma^{\mathrm{tr}}-\bchi_n):\boldsymbol{R}}
{c\bar{\omega}\,(\bsigma^{\mathrm{tr}}-\bchi_n):\bN^{\mathrm{tr}}},
\label{it_1}
\\
\boldsymbol {R}=\left[\omega'\,\boldsymbol{\sigma}_{+}^{\mathrm{tr}}+\omega\,\mathbb{C}\!:\![\frac{({\boldsymbol{\epsilon}}_{}-{\boldsymbol{\epsilon}}_{n})}{\Delta t}-\lambda_{i-1}\bN^{\mathrm{tr}}]_{+}
\right.
\left.
+ \,\mathbb{C}\!:\![\frac{({\boldsymbol{\epsilon}}_{}-{\boldsymbol{\epsilon}}_{n})}{\Delta t}-\lambda_{i-1}\bN^{\mathrm{tr}}]_{-}  
 +\dot{\bar{\omega}}\,c\boldsymbol{\xi}_n
+\frac {c}{b}\,\bchi_n\right].
\label{it_2}
\end{gather}
The previous expressions use the approximation $\dot{\boldsymbol{\epsilon}}=\frac{({\boldsymbol{\epsilon}}_{}-{\boldsymbol{\epsilon}}_{n})}{\Delta t}$ and notation $\bN^{\mathrm{tr}}= \frac{(\bsigma^\mathrm{tr}-\bchi^\mathrm{n})}{\lvert\lvert\bsigma^\mathrm{tr}-\bchi^\mathrm{n}\rvert\rvert}$. Note that index $i$ is related to the iterative solution  for $\lambda_i$, whereas index $n$ denotes the time step. Quantities without any index are related to the step $n+1$ or are constant.
Finally, the solution of the iterative procedure \eqref{it_1}-\eqref{it_2} is denoted $\lambda=\lambda_i$ and used   to update plastic deformations $\boldsymbol{\epsilon}^\mathrm{p}_{}$, driving forces $\boldsymbol{\chi}$ and stress response $\boldsymbol{\sigma}$ 

\begin{gather}
\dot{\boldsymbol{\epsilon}}_\mathrm{p}=\frac{\boldsymbol{\epsilon}^\mathrm{p}_{}-\boldsymbol{\epsilon}^\mathrm{p}_{n}}{\Delta\mathrm{t}} = \lambda \bN^{\mathrm{tr}}
\hspace{27.8mm}\qquad\Rightarrow\quad
 \boldsymbol{\epsilon}^\mathrm{p}_{} = \Delta t\lambda\bN^{\mathrm{tr}} 
+ \boldsymbol{\epsilon}^\mathrm{p}_{n}, 
\\
\dot{\boldsymbol{\xi}}=\frac{\boldsymbol{\xi}-\boldsymbol{\xi}_n}{\Delta t}=
\frac{1}{b\bar{\omega}}\bchi_n 
- \dot{\boldsymbol{\epsilon}}_\mathrm{p}
\hspace{22mm}\qquad\Rightarrow\quad
\boldsymbol{\xi}= \frac{\Delta t}{b\bar{\omega}}\bchi_n -\Delta t\,\lambda\bN^{\mathrm{tr}} +\boldsymbol{\xi}_n,
\\
\dot{\boldsymbol{\chi}}=\dot{\bar{\boldsymbol{\chi}}} = \frac{\bchi_{}-\bchi_{n}}{\Delta\mathrm{t}} 
=-\dot{\bar{\omega}}\,c\boldsymbol{\xi}_n
+ c{\bar{\omega}}\dot{\boldsymbol{\epsilon}}_\mathrm{p}
-\frac{c}{b} \, \boldsymbol{\chi}_{n} 
\quad\Rightarrow\quad
\boldsymbol{\chi}=\bchi=
-\Delta t\,\dot{\bar{\omega}}\,c\boldsymbol{\xi}_n
+ c \bar{\omega}\,\Delta t\,\lambda\bN^{\mathrm{tr}} 
- \frac{c \Delta t}{b}\bchi_{n}+\bchi_{n},
\\[1mm]
	\boldsymbol{\sigma}_\mathrm{} = \omega(d) \ \mathbb{C}\!:\!(\boldsymbol{\epsilon}_{}-\boldsymbol{\epsilon}^\mathrm{p}_{})_{+} +\mathbb{C}\!:\!(\boldsymbol{\epsilon}_{}-\boldsymbol{\epsilon}^\mathrm{p}_{})_{-}.
\end{gather}


\section{Representative numerical examples}

The model developed inherits the advantages of both methods that it incorporates. That makes its application field large, as demonstrated in the subsequent  sections studying the behavior of the cold-formed carbon steel (CS) and of the cold-formed stainless steel (SS). Among others,
selected numerical examples deal with the purely elasto-plastic material behavior (Sect. \ref{sec_plas}), with the  simulation of crack propagation on a notched sample (Sect. \ref{sec_notch}) and with  the study of the life time in the LCF-mode (Sect.\ref{sec_life}).

\subsection{Elasto-plastic behavior of the carbon steel and of the stainless steel}
\label{sec_plas}

The first group of examples illustrating the application of the model studies a purely elasto-plastic material behavior under a cyclic load by neglecting the damage effects. To this end, simulations at a single material point are performed for  parameter sets shown in Tab. \ref{el_pl}. The prescribed strains gradually change from zero to 1~\% and thereafter decrease to -1~\%. The cycle again closes  at zero load level. Two cycles with identical load paths are carried out. Each cycle takes 2 s to complete, whereas  a single time step is $\Delta t=0.01$ s.

\begin{table}
  \centering
  \caption{Elastic and plastic material parameters for CS and SS.}
  \label{el_pl}
  \begin{tabular}{lllll}
    \toprule
    	Parameter\hspace{5mm} & Value CS\hspace{5mm} & Value SS\hspace{5mm} & Unit\hspace{5mm} & Name \\
	\midrule
	$E$ & 212910& 197960 & [MPa] & Young's modulus \\
	$\nu$ & 0.3 & 0.3 & [--] & Poisson's ratio \\
	$\sigma_\mathrm{Y}$ & 451 & 552 & [MPa] & yield stress \\ 
	$b$ & 5 & 6 & [Ns/mm$^2$] \hspace{5mm} & pseudo-viscosity \\
	$c$ & 30000& 70000 & [N/mm] &   kinematic hardening parameter \\
    \bottomrule
  \end{tabular}
\end{table}

\begin{figure}[!ht]
	\centering
	\includegraphics[width = \textwidth]{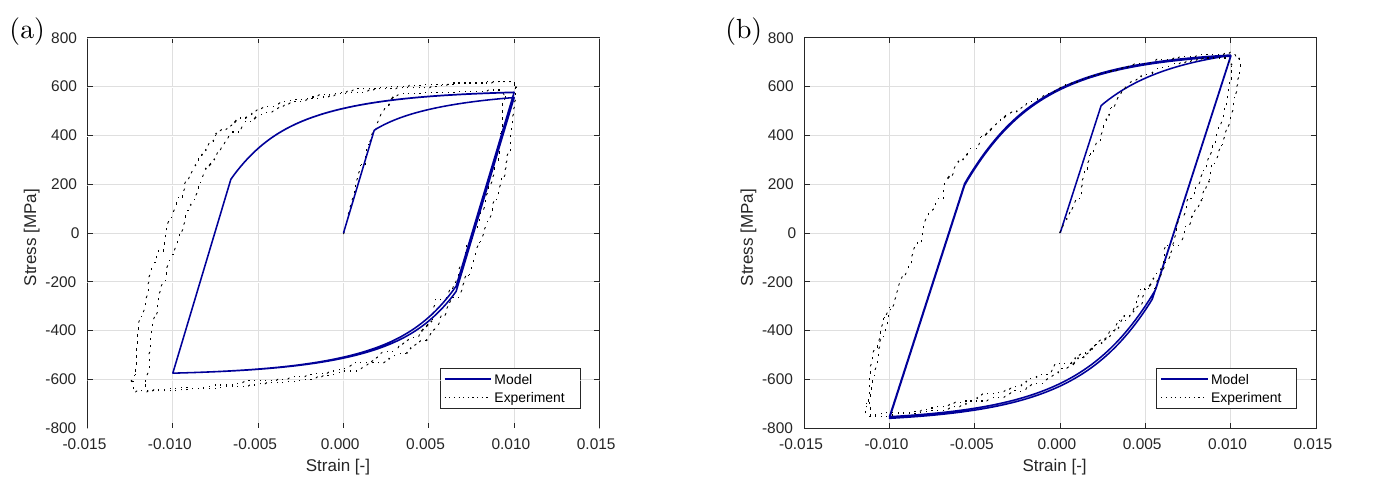}
	\caption{Comparison of experimental and numerical results at a material point. (a) Stress-strain hysteresis for carbon steel (CS). (b) Stress-strain hysteresis for stainless steel (SS). Plotted are the 11-components of the stress tensor and the strain tensor. Experimental results are taken from \cite{nip1}. }
	\label{stress_cs_ss}
\end{figure}

Simulations are carried out for two kinds of steel, namely for the cold-formed carbon steel (CS) and the cold-formed stainless steel (SS). The results are furthermore compared to the experimental findings by \cite{nip1}. This comparison is presented in Fig. \ref{stress_cs_ss} and approves an excellent agreement. The stainless steel hardens stronger  than the carbon steel, such that parameter $c$ is higher in this case.   
More details on the parameter identification in the context of  the Armstrong-Frederick model are  for example provided in \cite{Wolff10, Lubarda02}.

\subsection{Crack propagation on a notched sample}\label{sec_notch}

\begin{figure}[!ht]
	\centering
	\includegraphics[width = \textwidth]{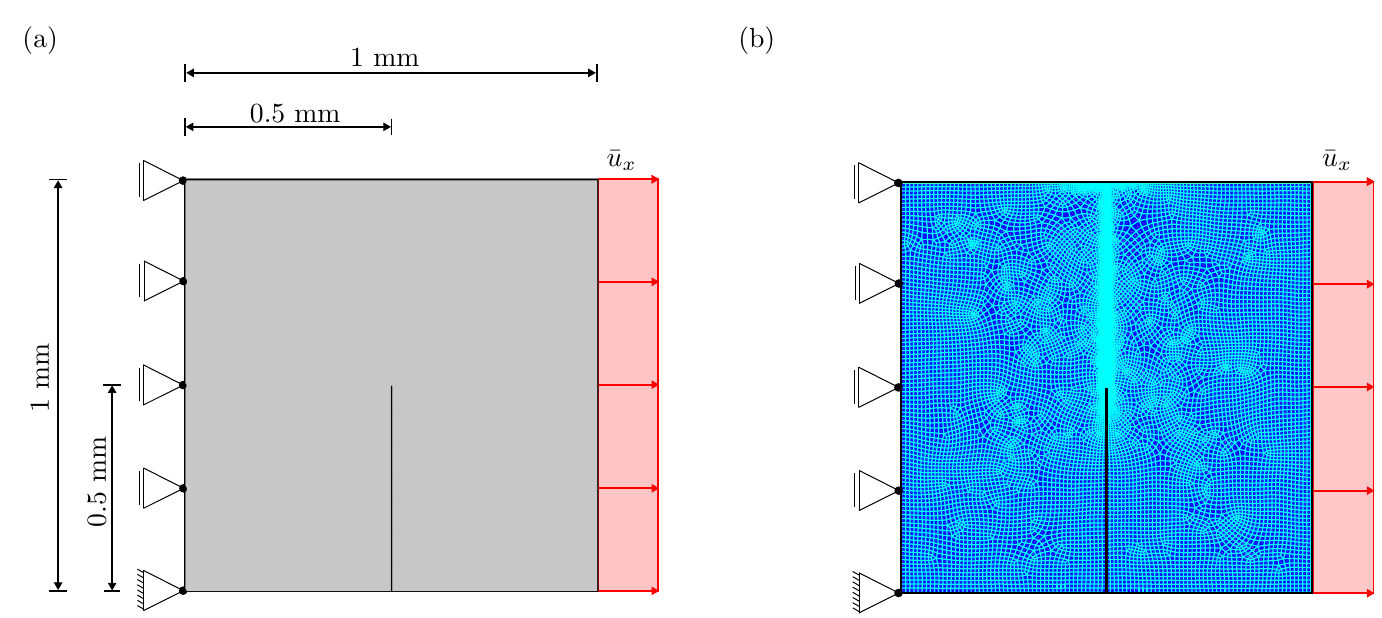}
	\caption{(a) Geometry of the notched plate with prescribed horizontal displacements $\bar{u}_x$. (b) Discretization of the plate with approximately 22500 elements.}
	\label{fig_setup}
\end{figure}

The  geometry of the sample for the second group of simulations is shown in Fig. \ref{fig_setup}. The chosen square plate has  dimensions 1 mm $\times$ 1 mm and a thickness of 1 mm and represents a  cutout of a  3D body with the large thickness, which substantiates the assumption of plane strains (Fig. \ref{fig_setup}a). The model does not include any material length scale, such that sample size and prescribed displacements can be scaled in a straightforward manner. The left edge of the sample is fixed in horizontal direction. The vertical displacement at the bottom left corner is additionally suppressed. Horizontal displacements $\bar{u}_x$ are prescribed at the right edge. The plate has a vertical initial crack from the middle of the bottom edge to the midpoint of the sample. It is discretized by a mesh with approximately 22500 quadrilateral elements such that a fine discretization is performed in the areas where crack propagation is expected (Fig. \ref{fig_setup}b). In this area, the effective element size $h$ fulfills the condition $h \approx 0.001 \text{ mm} < l/2$ and is significantly less than the minimum size required to achieve reasonable accuracy in the crack zone \citep{miehe2010thermodynamically}. Here, $l$ denotes the  characteristic crack width typical of the phase field method. 

The section simulates the crack propagation on the notched example and investigates the influence of material parameters on this process. The carbon steel is assumed as the representative material this time. Apart from elastic and plastic material parameters (Tab. \ref{el_pl}), simulations require damage parameters. The values chosen to this end are listed in Tab. \ref{dam}.
 Here, the values proposed by \cite{miehe2010thermodynamically} are used for the  critical energy $g_c$ and  characteristic crack width $l$. 
Constant $k_{\mathrm{p}}$ of the penalty function (Eq. \eqref{penalty}) is chosen to be $1 \times 10^8$, which is sufficiently large to suppress the negative damage rate. Constant $k_{\mathrm{d}}$ has the value of $1 \times 10^{-8}$ and prevents  the energy from becoming identical to zero in the case of full material damage.
In all examples of this group, the load increment and damage parameters $g_c$ and $l$ are assumed such that the crack formation and propagation can be simulated within a fairly small number of time steps in order to reduce the computational effort and to provide a qualitative analysis of relevant phenomena according to the results from \cite{miehe2010thermodynamically}.

\begin{table}
  \centering
  \caption{Material parameters of the damage model.}
  \label{dam}
  \begin{tabular}{lllll}
    \toprule
    	Parameter\hspace{5mm} & Value \hspace{5mm}  & Unit\hspace{5mm} & Name \\
	\midrule
	$g_c$ & 0.27 & [N/mm$^2$] & critical energy release rate \\
	$l$ & 0.0375  & [mm] & characteristic crack width \\
	$k_{\mathrm{p}}$  & $1 \times 10^{8}$ & [--] & penalty constant \\
	$k_{\mathrm{d}}$ & $1 \times 10^{-8}$ & [--] & numerical constant \\
    \bottomrule
  \end{tabular}
\end{table}

The behavior of the notched sample (Fig. \ref{fig_setup}) is investigated for three load constellations: uniformly increasing tension, cyclic load with a tension and compression phase and cyclic load with an increasing amplitude in the tension regime.

\begin{figure}[!ht]
	\centering
	\includegraphics[width = \textwidth]{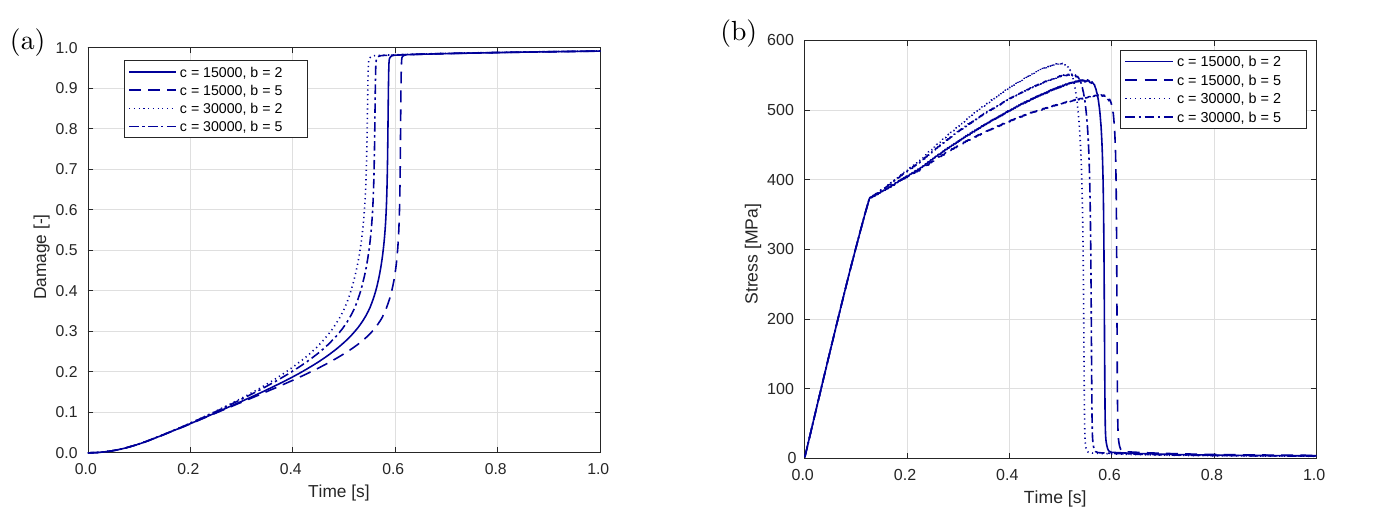}
	\caption{Results of tensile tests for different parameter pairs $b$ and $c$. Prescribed displacements uniformly change up to the value of 0.005 mm. (a) Applied load as a function of time and damage evolution over time.   (b) The 11-component of the stress tensor over the time. Hardening modulus $c$ is expressed in N/mm and pseudo-viscosity $b$ is expressed in Ns/mm$^2$.}
	\label{fig_tensile_test1}
\end{figure}

In the first case, the prescribed displacement linearly increases up to the maximum value of  $\bar{u}_x = 0.005$ mm. The constant time increment is  $\Delta t~=~0.005$~s and the total loading time amounts to 1 s. The results at a  quadrature point of an element directly located at the end of the initial crack are monitored for the illustration in Fig. \ref{fig_tensile_test1}. This figure shows the change of  damage and stress state during tension tests for different pairs of parameters $b$ and $c$.  
Here, rapidly hardening materials with a high hardening modulus $c$ lead to a faster increase of damage, whereas the higher pseudo-viscosity slows down the damage evolution and postpones softening. Damage increases from moderate values to the maximum, such that the material loses its strength and stress falls to the nearly zero value. The stress drop  takes place in approximately 0.05 s.  Simulations are performed for different  time increments in order to check the accuracy of the solution. For a time increment  $\Delta t~=~0.005$~s and less, identical results are achieved. 

\begin{figure}[!ht]
	\centering
	\includegraphics[width = \textwidth]{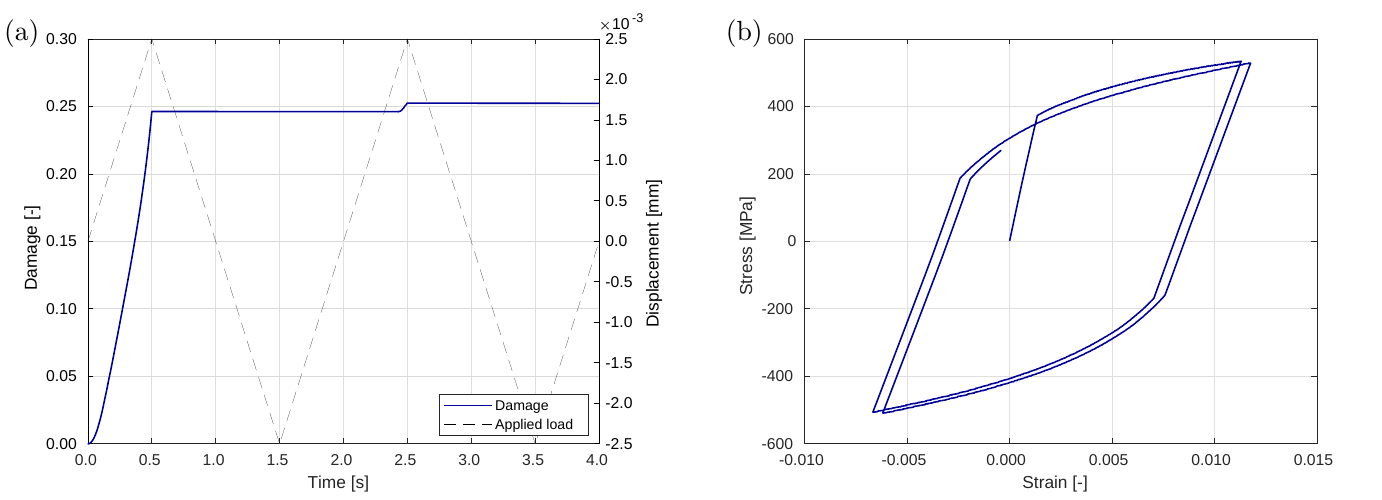}
	\caption{(a)   Applied load as a function of time and the damage evolution for two load cycles. (b) The 11-component of the stress tensor over the 11-component of the strain tensor.}
	\label{fig_cyclic_test1}
\end{figure}

The same setup (Fig. \ref{fig_setup}) is used to investigate influences of a cyclic load as shown in Fig. \ref{fig_cyclic_test1}.  Here,  two load cycles are performed with the total duration of 4 s. The time increment is $\Delta t = 0.005$ s. The horizontal displacement $\bar{u}_x$ changes in the range [0.0025 mm, -0.0025 mm], implying that a cycle includes both: the tension and the compression mode. The load amplitudes are constant in both cycles.  Figure \ref{fig_cyclic_test1}a  shows the load path and the damage evolution. The damage variable evolves during the tension phase, whereas the unloading and the compression mode do not affect it. This behavior goes back to the split of the energy into a tension and a compression part (Eq. \eqref{energysplit}). Even though the same displacement is applied in every cycle, an increase in damage is observed during each loading cycle. 
Figure \ref{fig_cyclic_test1}b shows the strong influence of damage on the corresponding stress-strain hysteresis. Here, the Baushinger effect is hardly noticeable, although a displacement controlled test with constant amplitudes is simulated.
 Fig. \ref{fig_cyclic_test1}b depicts the ratcheting rather than the Bauschinger effect.

\begin{figure}[!ht]
	\centering
	\includegraphics[width = 0.45\textwidth]{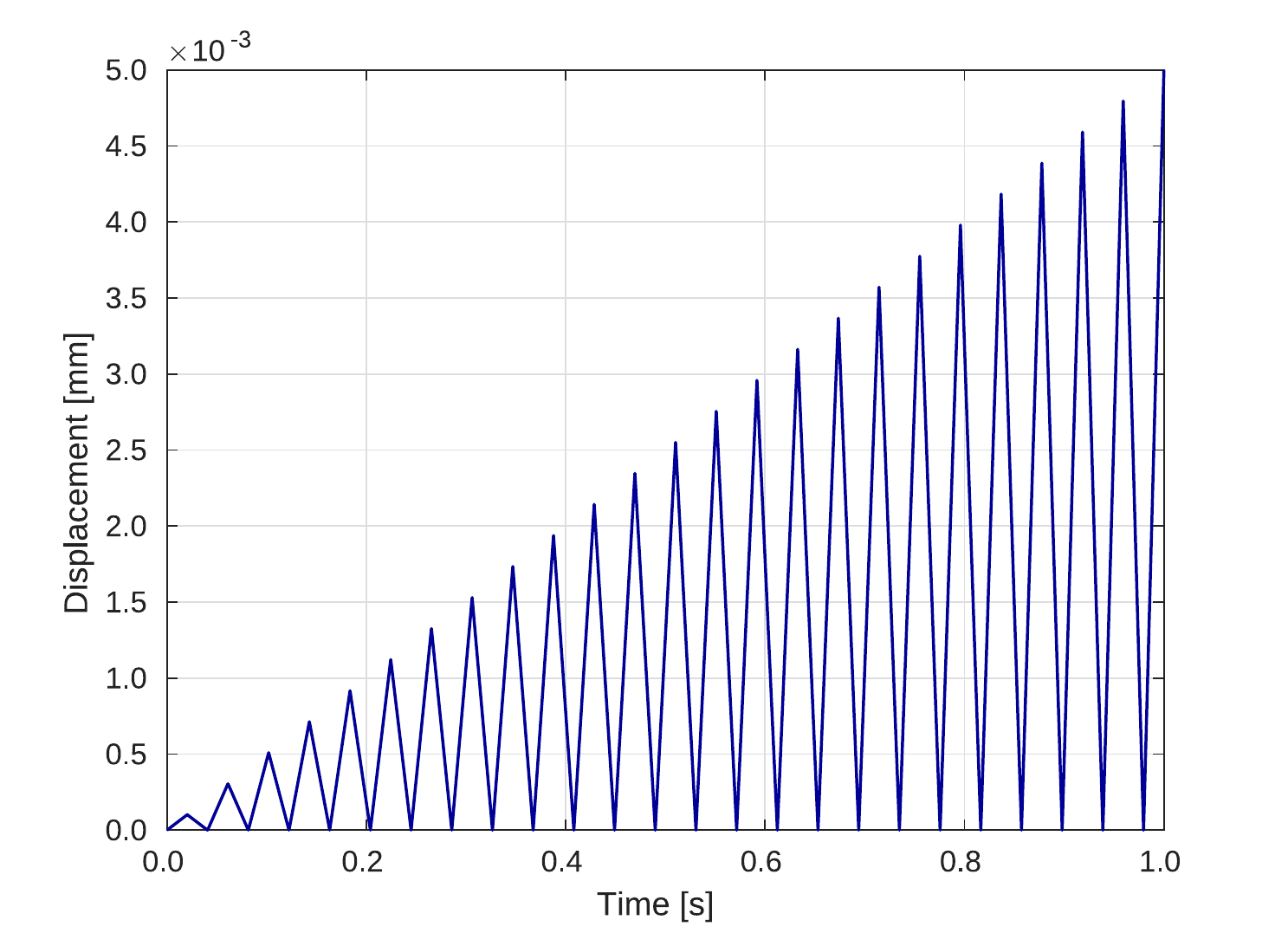}
	\caption{Increasing load amplitudes as a function of time. }
	\label{cyc_load}
\end{figure}

\begin{figure}[!ht]
	\centering
	\includegraphics[width = \textwidth]{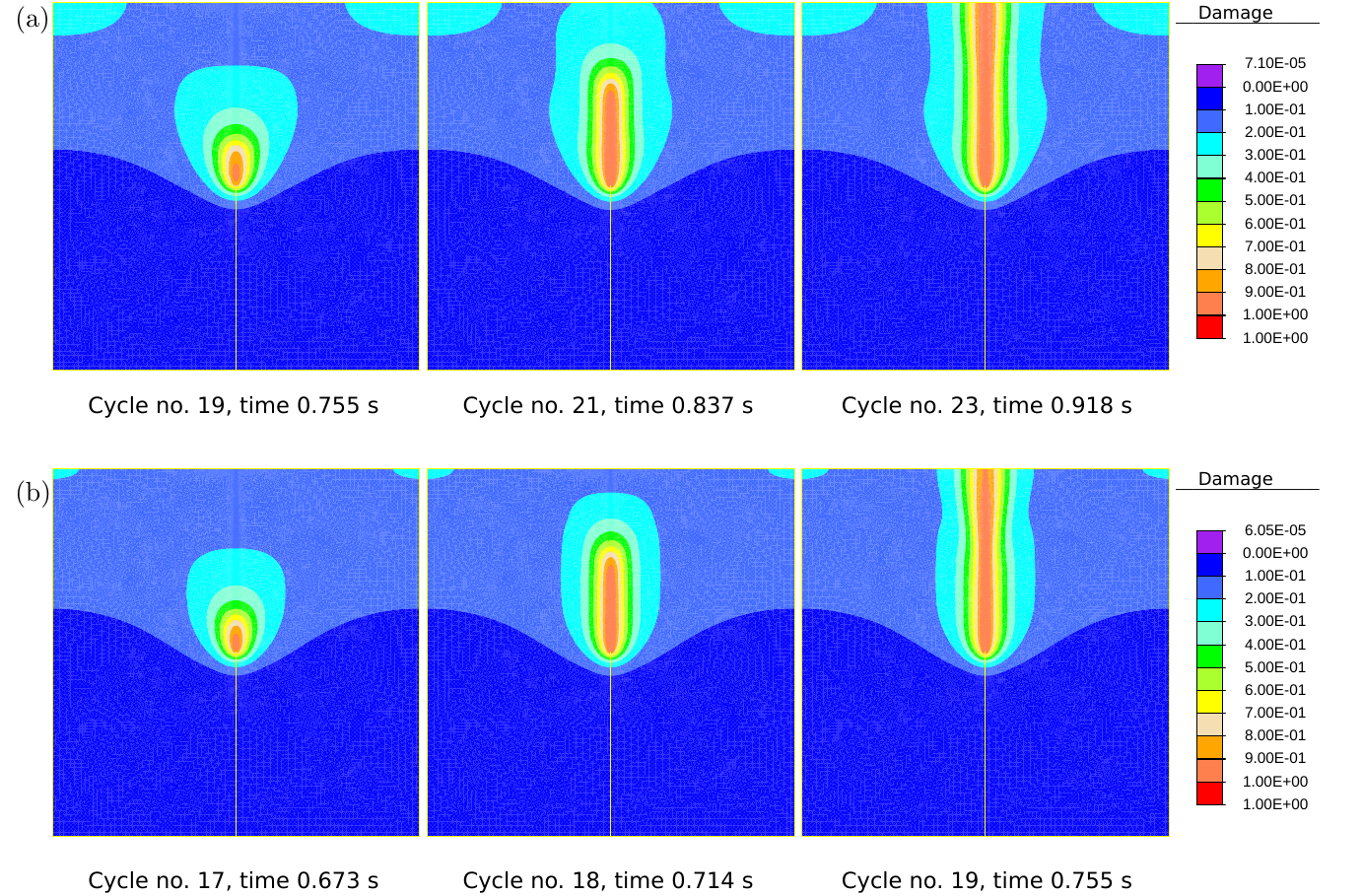}
	\caption{(a) Crack propagation takes place in five cycles ($c=15000$ N/mm, $b=2$ Ns/mm$^2$). (b) Crack propagation takes place in three cycles ($c=30000$ N/mm, $b=2$ Ns/mm$^2$).}
	\label{FE_res}
\end{figure}

 The last load constellation applied to the  setup from Fig. \ref{fig_setup} deals with the cyclic load in the tension regime with the increasing amplitude (Fig. \ref{cyc_load}).
 The amplitude increment amounts to $2 \times 10^{-4}$ mm per cycle. Time increment $\Delta t = 0.005$ s is assumed.
Three damage contour plots are chosen presenting states of the crack during propagation for parameters $c=15000$ N/mm (Fig. \ref{FE_res}a) and $c=30000$ N/mm (Fig. \ref{FE_res}b). The first snapshot indicates the state at which damage variable $d$ reaches a value of one at the already existing crack tip.
The second plot is taken when the crack has propagated roughly half the way through the plate, and the last plot is taken when the plate is fully torn in half.  For the first parameter value ($c=15000$ N/mm), the crack opening starts in the 19th load cycle and is fully propagated throughout the plate in the 23rd load cycle (Fig. \ref{FE_res}a).
For the second parameter value ($c=30000$ N/mm),  damage evolves earlier and faster: here, the complete
crack propagation takes place within three load cycles (Fig. \ref{FE_res}b).


\newpage

\subsection{Life time of the cold steel and of the stainless steel in the LCF-mode}
\label{sec_life}

Amongst others, the model proposed enables the estimation of the life time  in the LCF-mode. To this end, the number of cycles  up to the total failure is evaluated dependent on  the  amplitude of the cyclic load applied. The latter is kept  constant during the test. The material failure is caused by the material defects such as production process induced pores at a small length scale. In most cases, surface porosity is  the  critical factor for the fatigue phenomenon. However, the present paper assumes that different methods of the surface treatment can resolve this type of imperfection and that a defect in the bulk of material causes the crack initiation.

\begin{figure}[!ht]
	\centering
	\includegraphics[width = 0.9\textwidth]{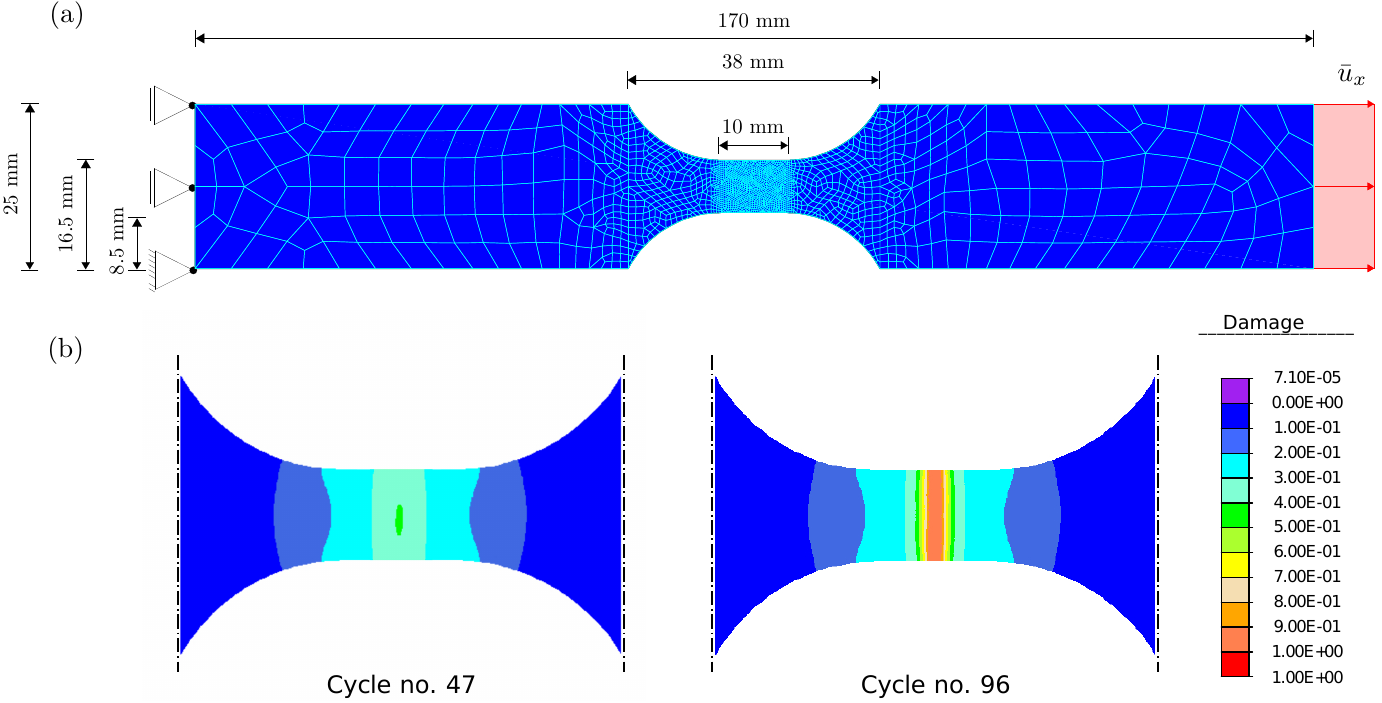}
	\vspace*{0mm}
	\caption{(a) Geometry of the simulated sample and prescribed boundary conditions. (b) Different stages of the crack propagation under the cyclic load. Simulations are performed for the carbon steel with the hardening modulus $c=30000$ N/mm.}
	\label{fig_cyclic_test5}
\end{figure}

This part of the analysis assumes the sample geometry (Fig. \ref{fig_cyclic_test5}a) proposed by \cite{nip1}. The horizontal displacements are constrained at the left boundary and the  cyclic horizontal displacements are prescribed at the right boundary.  In addition, the vertical displacement at the lower left corner is constrained to suppress the rigid body motion. For the chosen boundary condition, the largest deformations are expected in the center of the sample, which is discretized by a fine mesh. The effective element size in this area is h $\approx$ 0.25 mm, such that the crack width length $l=0.6$ is chosen. The critical energy is set to $g_c=100$ N/mm$^2$ which corresponds to a ductile material behavior, whereas the remaining material parameters are kept as in Sects. \ref{sec_plas} and \ref{sec_notch}. Simulations are additionally performed for higher values of the hardening modulus $c$ in order to study the influence of this parameter. A defect is created in an element in the center of the specimen, which initiates crack growth in the otherwise homogeneous material. In this element, the values of the Young's modulus and the yield stress are  set to 75\% of the actual material parameters.

The very first step of the analysis applies a strain of 3 \%. The assumed time increment is $\Delta t = 4. 28 \times 10^{-4}$ s and  2000 time steps simulate a single cycle. As expected, a vertical crack propagates through the middle of the sample, which is shown in Fig. \ref{fig_cyclic_test5}b. Here, an early stage of the crack is already noticeable in the 47th cycle, whereas the total failure occurs  in cycle number 96. Figure \ref{fig_cyclic_test5}b corresponds to the carbon steel with the hardening modulus $c=30000$ N/mm.

Furthermore, the tests are repeated by increasing the  load amplitude to the strain of 4 \%, 5 \%, 6 \% and 7 \%. The same setup (Fig. \ref{fig_cyclic_test5}a) is applied to this end. The number of  time steps within a cycle is kept constant (2000), however, the time increment varies in the range  from $\Delta t = 4.28 \times 10^{-4}$ s to $\Delta t = 1 \times 10^{-3}$ s. Accordingly, the duration of one cycle takes the values between 0.856~s and 2 s.

\begin{figure}[!ht]
	\centering
	\includegraphics[width = \textwidth]{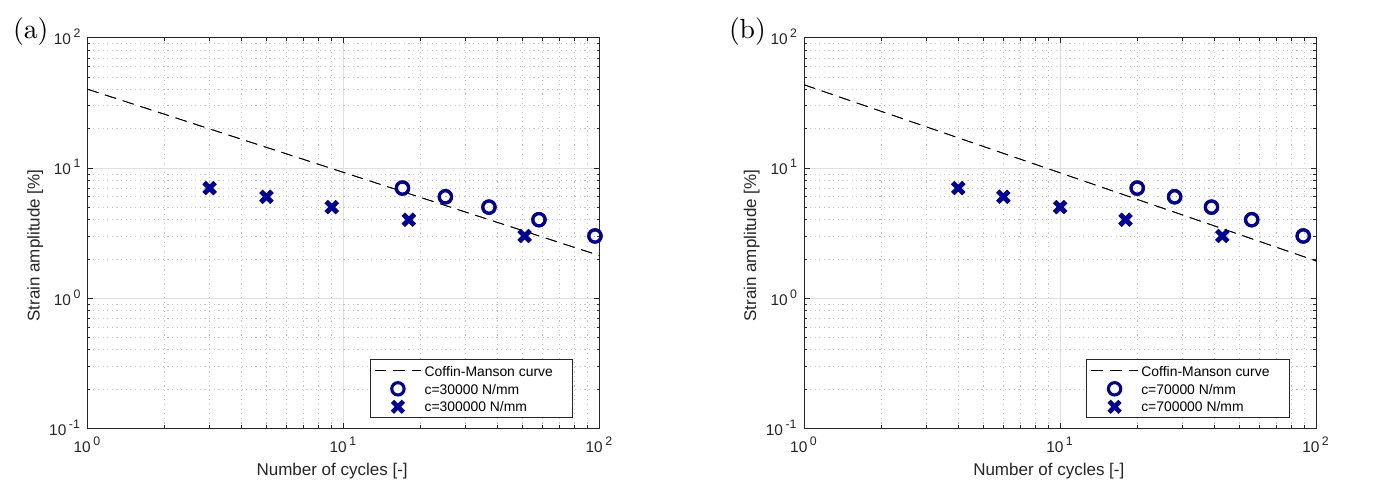}
	\caption{Comparison of the  experimental Coffin-Manson curve \citep{nip1} to the numerical results.   (a) Results for carbon steel with two different hardening parameters $c=30000$ N/mm and $c=300000$ N/mm. (b) Results  for the stainless steel with the hardening parameters $c=70000$ N/mm and $c=700000$ N/mm.}
	\label{fig_cyclic_test6}
\end{figure}

The results of  simulations together  with the experimental results by \cite{nip1} are presented in 
Fig. \ref{fig_cyclic_test6}. First, the behavior of carbon steel is studied and compared to the Coffin-Manson curve relating the applied strains to the number of cycles, both in logarithmic scales (Fig. \ref{fig_cyclic_test6}a). Numerical results show an excellent agreement with the experimental findings, in particular for  higher strains. The dependence between the applied strains and the number of cycles is linear, however, the number  of cycles is minimally overestimated. The discrepancy between the experimental and numerical results slightly increases with the decreasing strains. The same kind of simulations is repeated for a higher value  of the hardening parameter, which leads to a significant decrease of the number of  cycles up to the failure. This tendency goes back to the fact that a higher  hardening modulus is related  to the faster damage evolution and consequently causes a reduction of the number of cycles. Simulations for a higher hardening modulus approves  the linear  dependency between  data, however, they indicate that  a change of the strain amplitude  has a higher influence on the change of number of cycles in this case.
 
The observations mentioned previously have also been approved by repeating the same kind of tests for the stainless steel as shown in Fig. \ref{fig_cyclic_test6}b. Here, the Manson-Coffin curve  has a slightly higher slope, but numerical results  compared to the experimental ones show the same tendencies as for the carbon steel. 

\section{Conclusions and outlook}

The present work couples the phase field method of fracture to the Armstrong-Frederick model of plasticity with the kinematic hardening. The chosen concept inherits the advantages of both techniques and is aimed at the study of LCF effects in ductile materials.   However, the numerical implementation of this promising approach faces several challenges, such as the definition of a unique framework for both setups, the derivation of coupled evolution equations, the distinction between tension and compression mode and the development of a computationally efficient algorithm.  The basis for the phase field fracture model are the Griffith's theory and the dissipation potential relying on the assumption of a crack surface function. The derivation of evolution equations uses the minimum principle of the dissipation potential, which requires to express the dissipation potential of the classical Armstrong-Frederick model in terms of the internal variable rates by using the Legendre transformation.   The model developed also takes into account that the unloading and the compression mode have no effect on the damage evolution. The approach is eventually implemented in the FE-program FEAP where the displacements and damage are calculated at the global level by using a staggered scheme, whereas inelastic internal variables are updated locally, at each Gauss point.

The application of the model is illustrated by three groups of examples related to the cold-formed carbon steel and the cold-formed stainless steel. The first group of tests  investigates the  purely elasto-plastic material behavior without the influence of damage and focuses on the numerical reconstruction  of the stress-strain hysteresis due to a cyclic load. The second group of tests simulates the crack propagation on a notched sample and particularly analyses the influence of plastic parameters on the damage evolution. The simulations show  that the rapidly hardening materials are prone to a faster  damage evolution. The ability of the model to constrain the damage evolution  in the unloading and compression stages of a loading cycle  is  demonstrated as well. The last group of tests  studies the life time of the carbon steel in the LCF-mode and compares the numerical results to the experimental findings by \cite{nip1}. The results show an excellent qualitative and quantitative agreement, and the linear dependency  between the data in a logarithmic scale is approved.

The model proposed is a promising tool with regard to the  simulation of fatigue effects, giving rise to many new issues. Among others, it can be extended to capture ``hidden'' aspects of the fatigue process such as the microcrack initiation based on the theory of persistent slip bands and the microcrack propagation along the crystallographic planes. Moreover, the additional effects typical of high and very high cycle fatigue, as well as the threshold for the damage initiation have to be incorporated. A comprehensive  validation of the model with respect to the experimental results as well as its application and calibration for further materials are also envisaged.

\appendix 
\section{Legendre transformation}\label{appendix_a}
The formulation of the MDP in terms of rates of internal variables is based on Legendre transformation:

\begin{gather}
	\Phi^{\mathrm{AF}}(\boldsymbol{\nu},\dot{\boldsymbol{\nu}})=\underset{{\boldsymbol{q}}}{\text{max}}\{\mathcal{L}^\mathrm{LT}=\boldsymbol{q}:\dot{\boldsymbol{\nu}}-\Phi^{\mathrm{AF^*}}(\boldsymbol{\nu},{\boldsymbol{q}})\},
	\\
	\mathcal{L}^\mathrm{LT} = \boldsymbol{\sigma}:\dot{\boldsymbol{{\epsilon}}}^\mathrm{p}+\boldsymbol{\chi}:\dot{\boldsymbol{\xi}} - a ||\bsigma-\bchi||-\frac{1}{2b}||\bchi||^2. \label{a1}
\end{gather}
The corresponding stationary conditions have already been derived in Sect.\ref{sect_lengendre} (Eqs.\eqref{sig}b and\eqref{chii}b) 

\begin{equation}
	\bsigma=\byield\frac{\dot{\boldsymbol{{\epsilon}}}^\mathrm{p}}{||\dot{\boldsymbol{{\epsilon}}}^\mathrm{p}||}+b\dot{\boldsymbol{\xi}}+b\dot{\boldsymbol{{\epsilon}}}^\mathrm{p}, \qquad
\bchi=b\dot{\boldsymbol{\xi}}+b\dot{\boldsymbol{{\epsilon}}}^\mathrm{p}.
\end{equation}
Their implementation in \eqref{a1} now yields

\begin{gather}
\mathcal{L}^\mathrm{LT} = \boldsymbol{\sigma}:\dot{\boldsymbol{{\epsilon}}}^\mathrm{p}+\boldsymbol{\chi}:\dot{\boldsymbol{\xi}} - a ||\bsigma-\bchi||-\frac{1}{2b}||\bchi||^2= \notag\\
=\byield \frac{\dot{\boldsymbol{{\epsilon}}}^\mathrm{p}}{||\dot{\boldsymbol{{\epsilon}}}^\mathrm{p}||}:\dot{\boldsymbol{{\epsilon}}}^\mathrm{p}
+b \dot{\boldsymbol{\xi}}:\dot{\boldsymbol{{\epsilon}}}^\mathrm{p} 
+b \dot{\boldsymbol{{\epsilon}}}^\mathrm{p}:\dot{\boldsymbol{{\epsilon}}}^\mathrm{p}
+b \dot{\boldsymbol{\xi}}:\dot{\boldsymbol{\xi}}
+b \dot{\boldsymbol{{\epsilon}}}^\mathrm{p} :\dot{\boldsymbol{\xi}}
-a||\byield\frac{\dot{\boldsymbol{{\epsilon}}}^\mathrm{p}}{||\dot{\boldsymbol{{\epsilon}}}^\mathrm{p}||}|| \notag
-\frac{1}{2b}\left[ b^2 ||\dot{\boldsymbol{\xi}}||^2
+2b^2\dot{\boldsymbol{\xi}}:\dot{\boldsymbol{{\epsilon}}}^\mathrm{p}
+b^2 ||\dot{\boldsymbol{{\epsilon}}}^\mathrm{p}||^2 \right] \\
=\byield ||\dot{\boldsymbol{{\epsilon}}}^\mathrm{p}||
 +b \dot{\boldsymbol{\xi}}:\dot{\boldsymbol{{\epsilon}}}^\mathrm{p}
+b ||\dot{\boldsymbol{{\epsilon}}}^\mathrm{p}||^2
+b ||\dot{\boldsymbol{\xi}}||^2
+b \dot{\boldsymbol{\xi}}:\dot{\boldsymbol{{\epsilon}}}^\mathrm{p} 
-a\byield
-\frac{b}{2}||\dot{\boldsymbol{\xi}}||^2
-b \dot{\boldsymbol{\xi}}:\dot{\boldsymbol{{\epsilon}}}^\mathrm{p}
-\frac{b}{2} ||\dot{\boldsymbol{{\epsilon}}}^\mathrm{p}||^2 \notag\\
=\byield||\dot{\boldsymbol{{\epsilon}}}^\mathrm{p}|| 
+\frac{b}{2}||\dot{\boldsymbol{{\epsilon}}}^\mathrm{p}||^2+\frac{b}{2}||\dot{\boldsymbol{\xi}}||^2+b\,\dot{\boldsymbol{{\epsilon}}}^\mathrm{p}\!:\!\dot{\boldsymbol{\xi}} -a\byield.
\end{gather}
The last term is a constant and thus can be neglected, since the evolution equations are obtained by its minimization of the potential. Accordingly, the sought potential turns into

\begin{equation}
\Phi^{\mathrm{AF}}(\boldsymbol{\nu},\dot{\boldsymbol{\nu}})=\byield||\dot{\boldsymbol{{\epsilon}}}^\mathrm{p}||
+\frac{b}{2}||\dot{\boldsymbol{{\epsilon}}}^\mathrm{p}||^2+\frac{b}{2}||\dot{\boldsymbol{\xi}}||^2+b\,\dot{\boldsymbol{{\epsilon}}}^\mathrm{p}\!:\!\dot{\boldsymbol{\xi}}.
\end{equation}

\section{Evolution equations of the coupled problem}\label{appendix_b}

The potential of the coupled problem  has the form

\begin{equation} 
\mathcal{L}^{\mathrm{MDP}}=\dot{\Psi}^{\mathrm{c}}+\Phi^{\mathrm{c}}=
\dot{\Psi}^{\mathrm{c}}+\Phi^{\mathrm{d}}+\bar{\omega}\Phi^{\mathrm{AF}},
\end{equation}
where $\dot{\Psi}^{\mathrm{c}}$ is defined in Eq. \eqref{coupen}, $\Phi^{\mathrm{d}}$ is defined by Eq.  \eqref{dp_pen} and $\Phi^{\mathrm{AF}}$ by Eq. \eqref{klingepot}. Driving forces corresponding to ${\boldsymbol{\epsilon}^{\mathrm{p}}}$ and ${\boldsymbol{\xi}}$ are then calculated as following derivatives

\begin{gather}
\boldsymbol{q}_{\boldsymbol{\epsilon}^{\mathrm{p}}}=\bsigma=\frac{\partial \Phi^{\mathrm{c}}}{\partial{\dot{\boldsymbol{\epsilon}}^\mathrm{p}}}=
\bar{\omega}\left[
\byield\frac{\dot{\boldsymbol{{\epsilon}}}^\mathrm{p}}{||\dot{\boldsymbol{{\epsilon}}}^\mathrm{p}||}+ b{\dot{\boldsymbol{{\epsilon}}}^\mathrm{p}}+b{\dot{\boldsymbol{{\xi}}}}\right], \label{b_1}
\\
\boldsymbol{q}_{\boldsymbol{\xi}}=\bchi= \frac{\partial \Phi^{\mathrm{c}}}{\partial{\dot{\boldsymbol{\xi}}}}= \bar{\omega}\left[b{\dot{\boldsymbol{{\xi}}}}+b{\dot{\boldsymbol{{\epsilon}}}^\mathrm{p}}
\right] , \label{b_2}
\end{gather}
such that a transformation of the system \eqref{b_1} and \eqref{b_2} yields the evolution equations
 
\begin{equation}\label{ev_eq_one_}
{\dot{\boldsymbol{{\epsilon}}}^\mathrm{p}}=\lambda \frac{\bsigma-\bchi}{||\bsigma-\bchi||}, 
\qquad 
\dot{\boldsymbol{\xi}}=-{\dot{\boldsymbol{{\epsilon}}}^\mathrm{p}}+\frac{1}{\bar{\omega}b} \bchi,
\end{equation}
whereas  the insertion of \eqref{b_2} into \eqref{b_1} provides the  yield locus equation

\begin{equation}
\bsigma=\bar{\omega}\byield\frac{\dot{\boldsymbol{{\epsilon}}}^\mathrm{p}}{||\dot{\boldsymbol{{\epsilon}}}^\mathrm{p}||} + \boldsymbol{\chi}
\qquad\Rightarrow\qquad 
\bsigma-\bchi=\bar{\omega}\byield\frac{\dot{\boldsymbol{{\epsilon}}}^\mathrm{p}}{||\dot{\boldsymbol{{\epsilon}}}^\mathrm{p}||} 
\qquad\Rightarrow\qquad
||\bsigma-\bchi||=\bar{\omega}\byield. \label{b3}
\end{equation}
Finally, constitutive law \eqref{backstr}  is used to derive  the evolution equation for the back stress:

\begin{equation}
\dot{\boldsymbol{\chi}}=-\dot{\bar{\omega}}c \boldsymbol{\xi}-\bar{\omega}c\dot{\boldsymbol{\xi}}=-\dot{\bar{\omega}}c\boldsymbol{\xi}+ c\bar{\omega}\left(\dot{\boldsymbol{\epsilon}^\mathrm{p}}-\  \frac{1}{b\bar{\omega}} \  \bchi\right ).
\end{equation}

\vspace{6pt} 

\section*{Acknowledgments}
The authors are thankful to Frank Walther and Mustafa Awd (TU Dortmund University, Dortmund, Germany) for their helpful discussions on the fatigue behavior of metal materials.


\bibliographystyle{cas-model2-names}
\bibliography{literatureigen}

\begin{thebibliography}{48}
\expandafter\ifx\csname natexlab\endcsname\relax\def\natexlab#1{#1}\fi
\providecommand{\url}[1]{\texttt{#1}}
\providecommand{\href}[2]{#2}
\providecommand{\path}[1]{#1}
\providecommand{\DOIprefix}{doi:}
\providecommand{\ArXivprefix}{arXiv:}
\providecommand{\URLprefix}{URL: }
\providecommand{\Pubmedprefix}{pmid:}
\providecommand{\doi}[1]{\href{http://dx.doi.org/#1}{\path{#1}}}
\providecommand{\Pubmed}[1]{\href{pmid:#1}{\path{#1}}}
\providecommand{\bibinfo}[2]{#2}
\ifx\xfnm\relax \def\xfnm[#1]{\unskip,\space#1}\fi
\bibitem[{Abdollahi and Arias(2012)}]{ma1}
\bibinfo{author}{Abdollahi, A.}, \bibinfo{author}{Arias, I.},
  \bibinfo{year}{2012}.
\newblock \bibinfo{title}{Phase-field modeling of crack propagation in
  piezoelectric and ferroelectric materials with different electromechanical
  crack conditions}.
\newblock \bibinfo{journal}{J. Mech. Phys. Solids} \bibinfo{volume}{60},
  \bibinfo{pages}{2100--2126}.
\bibitem[{Alessi et~al.(2014)Alessi, Marigo and Vidoli}]{ma2}
\bibinfo{author}{Alessi, R.}, \bibinfo{author}{Marigo, J.J.},
  \bibinfo{author}{Vidoli, S.}, \bibinfo{year}{2014}.
\newblock \bibinfo{title}{Gradient damage models coupled with plasticity and
  nucleation of cohesive cracks}.
\newblock \bibinfo{journal}{Arch. Rat. Mech. Anal.} \bibinfo{volume}{214},
  \bibinfo{pages}{575--615}.
\bibitem[{Ambati et~al.(2015a)Ambati, Gerasimov and De~Lorenzis}]{ma3}
\bibinfo{author}{Ambati, M.}, \bibinfo{author}{Gerasimov, T.},
  \bibinfo{author}{De~Lorenzis, L.}, \bibinfo{year}{2015}a.
\newblock \bibinfo{title}{Phase-field modeling of ductile fracture}.
\newblock \bibinfo{journal}{Comput. Mech.} \bibinfo{volume}{55},
  \bibinfo{pages}{1017--1040}.
\bibitem[{Ambati et~al.(2015b)Ambati, Kruse and De~Lorenzis}]{ma4}
\bibinfo{author}{Ambati, M.}, \bibinfo{author}{Kruse, R.},
  \bibinfo{author}{De~Lorenzis, L.}, \bibinfo{year}{2015}b.
\newblock \bibinfo{title}{A phase-field model for ductile fracture at finite
  strains and its experimental verification}.
\newblock \bibinfo{journal}{Comput. Mech.} \bibinfo{volume}{57},
  \bibinfo{pages}{149--167}.
\bibitem[{Ambrosio and Tortorelli(1990)}]{ma5}
\bibinfo{author}{Ambrosio, L.}, \bibinfo{author}{Tortorelli, V.M.},
  \bibinfo{year}{1990}.
\newblock \bibinfo{title}{Approximation of functionals depending on jumps by
  elliptic functionals via t-convergence}.
\newblock \bibinfo{journal}{Commun- Pur. Appl. Math.} \bibinfo{volume}{43},
  \bibinfo{pages}{999--1036}.
\bibitem[{Ambrosio and Tortorelli(1992)}]{ma6}
\bibinfo{author}{Ambrosio, L.}, \bibinfo{author}{Tortorelli, V.M.},
  \bibinfo{year}{1992}.
\newblock \bibinfo{title}{On the approximation of free discontinuity problems}.
\newblock \bibinfo{journal}{Boll. Un. Mat. Ital. B(7)} \bibinfo{volume}{6},
  \bibinfo{pages}{105--123}.
\bibitem[{Armstrong and Frederick(1966)}]{armstrong1966mathematical}
\bibinfo{author}{Armstrong, P.J.}, \bibinfo{author}{Frederick, C.O.},
  \bibinfo{year}{1966}.
\newblock \bibinfo{title}{{A mathematical representation of the multiaxial
  Bauschinger effect}}.
\newblock \bibinfo{journal}{Central Electricity Generating Board and Berkeley
  Nuclear Laboratories} \bibinfo{volume}{731}.
\bibitem[{Aygün and Klinge(2020)}]{AYGUN2020}
\bibinfo{author}{Aygün, S.}, \bibinfo{author}{Klinge, S.},
  \bibinfo{year}{2020}.
\newblock \bibinfo{title}{Continuum mechanical modeling of strain-induced
  crystallization in polymers}.
\newblock \bibinfo{journal}{Int. J. Solids Struct.} \bibinfo{volume}{196--197},
  \bibinfo{pages}{129--139}.
\bibitem[{Bari and Hassan(2001)}]{r3}
\bibinfo{author}{Bari, S.}, \bibinfo{author}{Hassan, T.}, \bibinfo{year}{2001}.
\newblock \bibinfo{title}{Kinematic hardening rules in uncoupled modeling for
  multiaxial ratcheting simulation}.
\newblock \bibinfo{journal}{Int. J. Plast.} \bibinfo{volume}{17},
  \bibinfo{pages}{885--905}.
\bibitem[{Borden et~al.(2012)Borden, Verhoosel, Scott, Hughes and
  Landis}]{Borden12}
\bibinfo{author}{Borden, M.}, \bibinfo{author}{Verhoosel, C.},
  \bibinfo{author}{Scott, M.}, \bibinfo{author}{Hughes, T.},
  \bibinfo{author}{Landis, C.}, \bibinfo{year}{2012}.
\newblock \bibinfo{title}{A phase-field description of dynamic brittle
  fracture}.
\newblock \bibinfo{journal}{Comput. Methods Appl. Mech. Eng.}
  \bibinfo{volume}{217--220}, \bibinfo{pages}{77--95}.
\bibitem[{Borden et~al.(2016)Borden, Hughes, Landis, Anvari and Lee}]{ma14}
\bibinfo{author}{Borden, M.J.}, \bibinfo{author}{Hughes, T.J.R.},
  \bibinfo{author}{Landis, C.M.}, \bibinfo{author}{Anvari, A.},
  \bibinfo{author}{Lee, I.J.}, \bibinfo{year}{2016}.
\newblock \bibinfo{title}{A phase-field formulation for fracture in ductile
  materials: Finite deformation balance law derivation, plastic degradation,
  and stress triaxiality effects}.
\newblock \bibinfo{journal}{Comput. Methods Appl. Mech. Eng.}
  \bibinfo{volume}{312}, \bibinfo{pages}{130--166}.
\bibitem[{Bourdin and Chambolle(2000)}]{ma18}
\bibinfo{author}{Bourdin, B.}, \bibinfo{author}{Chambolle, A.},
  \bibinfo{year}{2000}.
\newblock \bibinfo{title}{Implementation of an adaptive finite-element
  approximation of the mumford-shah functional}.
\newblock \bibinfo{journal}{Numer. Math.} \bibinfo{volume}{85},
  \bibinfo{pages}{609--646}.
\bibitem[{Bourdin et~al.(2012)Bourdin, Chukwudozie and Yoshioka}]{ma19}
\bibinfo{author}{Bourdin, B.}, \bibinfo{author}{Chukwudozie, C.},
  \bibinfo{author}{Yoshioka, K.}, \bibinfo{year}{2012}.
\newblock \bibinfo{title}{A variational approach to the numerical simulation of
  hydraulic fracturing}, in: \bibinfo{booktitle}{SPE Annual Technical
  Conference and Exhibition}, \bibinfo{organization}{Society of Petroleum
  Engineers}. pp. \bibinfo{pages}{1--9}.
\bibitem[{Bourdin et~al.(2000)Bourdin, Francfort and Marigo}]{ma20}
\bibinfo{author}{Bourdin, B.}, \bibinfo{author}{Francfort, G.A.},
  \bibinfo{author}{Marigo, J.J.}, \bibinfo{year}{2000}.
\newblock \bibinfo{title}{Numerical experiments in revisited brittle fracture}.
\newblock \bibinfo{journal}{J. Mech. Phys. Solids} \bibinfo{volume}{48},
  \bibinfo{pages}{797--826}.
\bibitem[{Bourdin et~al.(2014)Bourdin, Marigo, Maurini and Sicsic}]{ma22}
\bibinfo{author}{Bourdin, B.}, \bibinfo{author}{Marigo, J.J.},
  \bibinfo{author}{Maurini, C.}, \bibinfo{author}{Sicsic, P.},
  \bibinfo{year}{2014}.
\newblock \bibinfo{title}{Morphogenesis and propagation of complex cracks
  induced by thermal shocks}.
\newblock \bibinfo{journal}{Phys. Rev. Lett.} \bibinfo{volume}{112},
  \bibinfo{pages}{014301}.
\bibitem[{Crismale and Lazzaroni(2016)}]{ma33}
\bibinfo{author}{Crismale, V.}, \bibinfo{author}{Lazzaroni, G.},
  \bibinfo{year}{2016}.
\newblock \bibinfo{title}{Viscous approximation of quasistatic evolutions for a
  coupled elastoplastic-damage model}.
\newblock \bibinfo{journal}{Calc. Var. Partial Dif.} \bibinfo{volume}{55},
  \bibinfo{pages}{1--54}.
\bibitem[{Dettmer and Reese(2004)}]{dettmer2004theo}
\bibinfo{author}{Dettmer, W.}, \bibinfo{author}{Reese, S.},
  \bibinfo{year}{2004}.
\newblock \bibinfo{title}{On the theoretical and numerical modelling of
  {Armstrong-Frederick} kinematic hardening in the finite strain regime}.
\newblock \bibinfo{journal}{Comput. Methods Appl. Mech. Eng.}
  \bibinfo{volume}{193}, \bibinfo{pages}{87--116}.
\bibitem[{Francfort and Marigo(1998)}]{ma40}
\bibinfo{author}{Francfort, G.A.}, \bibinfo{author}{Marigo, J.J.},
  \bibinfo{year}{1998}.
\newblock \bibinfo{title}{Revisiting brittle fracture as an energy minimization
  problem}.
\newblock \bibinfo{journal}{J. Mech. Phys. Solids.} \bibinfo{volume}{46},
  \bibinfo{pages}{1319--1342}.
\bibitem[{Fraternali(2007)}]{ma41}
\bibinfo{author}{Fraternali, F.}, \bibinfo{year}{2007}.
\newblock \bibinfo{title}{Free discontinuity finite element models in
  two-dimensions for in-plane crack problems}.
\newblock \bibinfo{journal}{Theor. Appl. Fract. Mec.} \bibinfo{volume}{47},
  \bibinfo{pages}{274--282}.
\bibitem[{Freddi and Iurlano(2017)}]{ma42}
\bibinfo{author}{Freddi, F.}, \bibinfo{author}{Iurlano, F.},
  \bibinfo{year}{2017}.
\newblock \bibinfo{title}{Numerical insight of a variational smeared approach
  to cohesive fracture}.
\newblock \bibinfo{journal}{J. Mech. Phys. Solids} \bibinfo{volume}{98},
  \bibinfo{pages}{156--171}.
\bibitem[{Hofacker and Miehe(2013)}]{Hofacker13}
\bibinfo{author}{Hofacker, M.}, \bibinfo{author}{Miehe, C.},
  \bibinfo{year}{2013}.
\newblock \bibinfo{title}{A phase field model of dynamic fracture: Robust field
  updates for the analysis of complex crack patterns}.
\newblock \bibinfo{journal}{Int. J. Numer. Methods Eng.} \bibinfo{volume}{93},
  \bibinfo{pages}{276--301}.
\bibitem[{Karma et~al.(2001)Karma, Kessler and Levine}]{ma57}
\bibinfo{author}{Karma, A.}, \bibinfo{author}{Kessler, D.A.},
  \bibinfo{author}{Levine, H.}, \bibinfo{year}{2001}.
\newblock \bibinfo{title}{{Phase-field model of mode III dynamic fracture}}.
\newblock \bibinfo{journal}{Phys. Rev. Lett.} \bibinfo{volume}{87},
  \bibinfo{pages}{045501}.
\bibitem[{Khan and Jackson(1999)}]{r15}
\bibinfo{author}{Khan, A.S.}, \bibinfo{author}{Jackson, K.M.},
  \bibinfo{year}{1999}.
\newblock \bibinfo{title}{{On the evolution of isotropic and kinematic
  hardening with finite plastic deformation, Part I: Compression/tension
  loading of OFHC copper cylinders}}.
\newblock \bibinfo{journal}{Int. J. Plast.} \bibinfo{volume}{15},
  \bibinfo{pages}{1265--1275}.
\bibitem[{Kobayashi and Ohno(2002)}]{r16}
\bibinfo{author}{Kobayashi, M.}, \bibinfo{author}{Ohno, N.},
  \bibinfo{year}{2002}.
\newblock \bibinfo{title}{Implementation of cyclic plasticity models based on a
  general form of kinematic hardening}.
\newblock \bibinfo{journal}{Int. J. Numer. Meth. Eng.} \bibinfo{volume}{53},
  \bibinfo{pages}{2217--2238}.
\bibitem[{Kuhn and Müller(2010)}]{Kuhn10}
\bibinfo{author}{Kuhn, C.}, \bibinfo{author}{Müller, R.},
  \bibinfo{year}{2010}.
\newblock \bibinfo{title}{A continuum phase field model for fracture}.
\newblock \bibinfo{journal}{Eng. Fract. Mech.} \bibinfo{volume}{77},
  \bibinfo{pages}{3625--3634}.
\newblock \bibinfo{note}{Computational Mechanics in Fracture and Damage: A
  Special Issue in Honor of Prof. Gross}.
\bibitem[{Li et~al.(2014)Li, Peco, Millan, Arias and Arroyo}]{ma63}
\bibinfo{author}{Li, B.}, \bibinfo{author}{Peco, C.}, \bibinfo{author}{Millan,
  D.}, \bibinfo{author}{Arias, I.}, \bibinfo{author}{Arroyo, M.},
  \bibinfo{year}{2014}.
\newblock \bibinfo{title}{Phase-field modeling and simulation of fracture in
  brittle materials with strongly anisotropic surface energy}.
\newblock \bibinfo{journal}{Int. J. Numer. Meth. Eng.} \bibinfo{volume}{102},
  \bibinfo{pages}{711--727}.
\bibitem[{Lion(2000)}]{r18}
\bibinfo{author}{Lion, A.}, \bibinfo{year}{2000}.
\newblock \bibinfo{title}{Constitutive modelling in finite
  thermoviscoplasticity: A physical approach based on nonlinear rheological
  models}.
\newblock \bibinfo{journal}{Int. J. Plast.} \bibinfo{volume}{16},
  \bibinfo{pages}{469--494}.
\bibitem[{Lubarda and Benson(2002)}]{Lubarda02}
\bibinfo{author}{Lubarda, V.}, \bibinfo{author}{Benson, D.},
  \bibinfo{year}{2002}.
\newblock \bibinfo{title}{On the numerical algorithm for isotropic–kinematic
  hardening with the {A}rmstrong–{F}rederick evolution of the back stress}.
\newblock \bibinfo{journal}{Comput. Methods Appl. Mech. Eng.}
  \bibinfo{volume}{191}, \bibinfo{pages}{3583--3596}.
\bibitem[{Lührs et~al.(1997)Lührs, Hartmann and Haupt}]{r19}
\bibinfo{author}{Lührs, G.}, \bibinfo{author}{Hartmann, S.},
  \bibinfo{author}{Haupt, P.}, \bibinfo{year}{1997}.
\newblock \bibinfo{title}{On the numerical treatment of finite deformations in
  elastoviscoplasticity}.
\newblock \bibinfo{journal}{Comput. Methods Appl. Mech. Eng.}
  \bibinfo{volume}{144}, \bibinfo{pages}{1--21}.
\bibitem[{Maurini et~al.(2013)Maurini, Bourdin, Gauthier and V.}]{ma68}
\bibinfo{author}{Maurini, C.}, \bibinfo{author}{Bourdin, B.},
  \bibinfo{author}{Gauthier, G.}, \bibinfo{author}{V., L.},
  \bibinfo{year}{2013}.
\newblock \bibinfo{title}{Crack patterns obtained by unidirectional drying of a
  colloidal suspension in a capillary tube: Experiments and numerical
  simulations using a two-dimensional variational approach}.
\newblock \bibinfo{journal}{Int. J. Fracture} \bibinfo{volume}{184},
  \bibinfo{pages}{75--91}.
\bibitem[{Miehe et~al.(2016)Miehe, Aldakheel and Raina}]{ma70}
\bibinfo{author}{Miehe, C.}, \bibinfo{author}{Aldakheel, F.},
  \bibinfo{author}{Raina, A.}, \bibinfo{year}{2016}.
\newblock \bibinfo{title}{{Phase field modeling of ductile fracture at finite
  strains: A variational gradient-extended plasticity-damage theory}}.
\newblock \bibinfo{journal}{Int. J. Plast.} \bibinfo{volume}{84},
  \bibinfo{pages}{1--32}.
\bibitem[{Miehe et~al.(2015)Miehe, Hofacker, Schänzel and Aldakheel}]{ma71}
\bibinfo{author}{Miehe, C.}, \bibinfo{author}{Hofacker, M.},
  \bibinfo{author}{Schänzel, L.M.}, \bibinfo{author}{Aldakheel, F.},
  \bibinfo{year}{2015}.
\newblock \bibinfo{title}{{Phase field modeling of fracture in multi-physics
  problems. Part II. Coupled brittle-to-ductile failure criteria and crack
  propagation in thermo-elastic-plastic solids}}.
\newblock \bibinfo{journal}{Comput. Methods Appl. Mech. Eng.}
  \bibinfo{volume}{294}, \bibinfo{pages}{486--522}.
\bibitem[{Miehe et~al.(2010a)Miehe, Hofacker and Welschinger}]{miehe2010phase}
\bibinfo{author}{Miehe, C.}, \bibinfo{author}{Hofacker, M.},
  \bibinfo{author}{Welschinger, F.}, \bibinfo{year}{2010}a.
\newblock \bibinfo{title}{A phase field model for rate-independent crack
  propagation: Robust algorithmic implementation based on operator splits}.
\newblock \bibinfo{journal}{Comput. Methods Appl. Mech. Eng.}
  \bibinfo{volume}{199}, \bibinfo{pages}{2765--2778}.
\bibitem[{Miehe et~al.(2010b)Miehe, Welschinger and
  Hofacker}]{miehe2010thermodynamically}
\bibinfo{author}{Miehe, C.}, \bibinfo{author}{Welschinger, F.},
  \bibinfo{author}{Hofacker, M.}, \bibinfo{year}{2010}b.
\newblock \bibinfo{title}{Thermodynamically consistent phase-field models of
  fracture: Variational principles and multi-field fe implementations}.
\newblock \bibinfo{journal}{Int. J. Numer. Meth. Eng.} \bibinfo{volume}{83},
  \bibinfo{pages}{1273--1311}.
\bibitem[{Mollica et~al.(2001)Mollica, Rajagopal and Srinivasa}]{r23}
\bibinfo{author}{Mollica, F.}, \bibinfo{author}{Rajagopal, K.R.},
  \bibinfo{author}{Srinivasa, A.R.}, \bibinfo{year}{2001}.
\newblock \bibinfo{title}{The inelastic behavior of metals subject to loading
  reversal}.
\newblock \bibinfo{journal}{Int. J. Plast.} \bibinfo{volume}{17},
  \bibinfo{pages}{1119--1146}.
\bibitem[{Msekh et~al.(2015)Msekh, Sargado, Jamshidian, Areias and
  Rabczuk}]{Msekh15}
\bibinfo{author}{Msekh, M.}, \bibinfo{author}{Sargado, J.M.},
  \bibinfo{author}{Jamshidian, M.}, \bibinfo{author}{Areias, P.},
  \bibinfo{author}{Rabczuk, T.}, \bibinfo{year}{2015}.
\newblock \bibinfo{title}{Abaqus implementation of phase-field model for
  brittle fracture}.
\newblock \bibinfo{journal}{Comput. Mater. Sci.} \bibinfo{volume}{96},
  \bibinfo{pages}{472--484}.
\bibitem[{Narayan and Anand(2019)}]{ma92}
\bibinfo{author}{Narayan, S.}, \bibinfo{author}{Anand, L.},
  \bibinfo{year}{2019}.
\newblock \bibinfo{title}{A gradient damage theory of fracture of quasi-brittle
  materials}.
\newblock \bibinfo{journal}{J. Mech. Phys. Solids} \bibinfo{volume}{129},
  \bibinfo{pages}{119--146}.
\bibitem[{Negri and Paolini(2001)}]{ma74}
\bibinfo{author}{Negri, M.}, \bibinfo{author}{Paolini, M.},
  \bibinfo{year}{2001}.
\newblock \bibinfo{title}{Numerical minimization of the mumford-shah
  functional}.
\newblock \bibinfo{journal}{Calcolo} \bibinfo{volume}{38},
  \bibinfo{pages}{67--84}.
\bibitem[{Nip et~al.(2010)Nip, Gardner, Davies and Elghazouli}]{nip1}
\bibinfo{author}{Nip, K.H.}, \bibinfo{author}{Gardner, L.},
  \bibinfo{author}{Davies, C.M.}, \bibinfo{author}{Elghazouli, A.Y.},
  \bibinfo{year}{2010}.
\newblock \bibinfo{title}{Extremely low cycle fatigue tests on structural
  carbon steel and stainless steel}.
\newblock \bibinfo{journal}{J. Constr. Steel Res.} \bibinfo{volume}{66},
  \bibinfo{pages}{96--110}.
\bibitem[{Ohno and Wang(1993)}]{r24}
\bibinfo{author}{Ohno, N.}, \bibinfo{author}{Wang, J.D.}, \bibinfo{year}{1993}.
\newblock \bibinfo{title}{{Kinematic hardening rules with critical state of
  dynamic recovery: Part I: Formulation and basic features for ratcheting
  behavior, Part II: Application to experiments of ratcheting behavior}}.
\newblock \bibinfo{journal}{Int. J. Plast.} \bibinfo{volume}{9},
  \bibinfo{pages}{375--403}.
\bibitem[{Puzrin and Houlsby(2001)}]{r25}
\bibinfo{author}{Puzrin, A.M.}, \bibinfo{author}{Houlsby, G.T.},
  \bibinfo{year}{2001}.
\newblock \bibinfo{title}{Fundamentals of kinematic hardening hyperplasticity}.
\newblock \bibinfo{journal}{Int. J. Solids Struct.} \bibinfo{volume}{38},
  \bibinfo{pages}{3771--3794}.
\bibitem[{Schlueter et~al.(2014)Schlueter, Willenbücher, Kuhn and
  Müller}]{Schlueter14}
\bibinfo{author}{Schlueter, A.}, \bibinfo{author}{Willenbücher, A.},
  \bibinfo{author}{Kuhn, C.}, \bibinfo{author}{Müller, R.},
  \bibinfo{year}{2014}.
\newblock \bibinfo{title}{Phase field approximation of dynamic brittle
  fracture}.
\newblock \bibinfo{journal}{Comput. Mech.} \bibinfo{volume}{54},
  \bibinfo{pages}{1141--1161}.
\bibitem[{Schmidt et~al.(2009)Schmidt, Fraternali and Ortiz}]{ma84}
\bibinfo{author}{Schmidt, B.}, \bibinfo{author}{Fraternali, F.},
  \bibinfo{author}{Ortiz, M.}, \bibinfo{year}{2009}.
\newblock \bibinfo{title}{Eigenfracture: An eigendeformation approach to
  variational fracture}.
\newblock \bibinfo{journal}{Multiscale Model. Sim.} \bibinfo{volume}{7},
  \bibinfo{pages}{1237--1266}.
\bibitem[{Svendsen et~al.(1998)Svendsen, Arndt, Klingbeil and Sievert}]{r38}
\bibinfo{author}{Svendsen, B.}, \bibinfo{author}{Arndt, S.},
  \bibinfo{author}{Klingbeil, D.}, \bibinfo{author}{Sievert, R.},
  \bibinfo{year}{1998}.
\newblock \bibinfo{title}{Hyperelastic models for elastoplasticity with
  nonlinear isotropic and kinematic hardening at large deformation}.
\newblock \bibinfo{journal}{Int. J. Solids Struct.} \bibinfo{volume}{35},
  \bibinfo{pages}{3363--3389}.
\bibitem[{Wheeler et~al.(2014)Wheeler, Wick and Wollner}]{ma89}
\bibinfo{author}{Wheeler, M.F.}, \bibinfo{author}{Wick, T.},
  \bibinfo{author}{Wollner, W.}, \bibinfo{year}{2014}.
\newblock \bibinfo{title}{An augmented-lagrangian method for the phase-field
  approach for pressurized fractures}.
\newblock \bibinfo{journal}{Comput. Methods Appl. Mech. Eng.}
  \bibinfo{volume}{271}, \bibinfo{pages}{69--85}.
\bibitem[{Wilson et~al.(2013)Wilson, Borden and Landis}]{ma90}
\bibinfo{author}{Wilson, Z.A.}, \bibinfo{author}{Borden, M.J.},
  \bibinfo{author}{Landis, C.M.}, \bibinfo{year}{2013}.
\newblock \bibinfo{title}{A phase-field model for fracture in piezoelectric
  ceramics}.
\newblock \bibinfo{journal}{Int. J. Fracture} \bibinfo{volume}{193},
  \bibinfo{pages}{135--153}.
\bibitem[{Wilson and Landis(2016)}]{ma91}
\bibinfo{author}{Wilson, Z.A.}, \bibinfo{author}{Landis, C.M.},
  \bibinfo{year}{2016}.
\newblock \bibinfo{title}{Phase-field modeling of hydraulic fracture}.
\newblock \bibinfo{journal}{J. Mech. Phys. Solids} \bibinfo{volume}{96},
  \bibinfo{pages}{264--290}.
\bibitem[{Wolff et~al.(2010)Wolff, Suhr and Simsir}]{Wolff10}
\bibinfo{author}{Wolff, M.}, \bibinfo{author}{Suhr, B.},
  \bibinfo{author}{Simsir, C.}, \bibinfo{year}{2010}.
\newblock \bibinfo{title}{Parameter identification for an
  {A}rmstrong–{F}rederick hardening law for supercooled austenite of {SAE}
  52100 steel}.
\newblock \bibinfo{journal}{Comput. Mater. Sci.} \bibinfo{volume}{50},
  \bibinfo{pages}{487--495}.

\end{thebibliography}

\end{document}